\PassOptionsToPackage{dvipsnames}{xcolor} 
\documentclass[twocolumn]{aastex631}

\usepackage{graphicx}	          
\usepackage{amsmath}	          
\usepackage{upgreek}            
\usepackage{amssymb}	          
\usepackage{bm}		              
\usepackage{CJKutf8}            
\usepackage[T1]{fontenc}        
\usepackage{comment}            
\usepackage{tabularx}           
\usepackage{threeparttable}     
\usepackage{booktabs}           
\usepackage{soul}               
\usepackage{ulem}               
\usepackage{multirow}           
\usepackage{microtype}          
\DisableLigatures[f]{encoding = *, family = *} 
\usepackage[figuresleft]{rotating}  
\allowdisplaybreaks             
\emergencystretch=\maxdimen     
\def\numx#1e#2{{#1}\mathrm{e}{#2}}
\newcommand{\mm}[1]{\mathrm{#1}}

\newcommand{\fracd}[2]{\frac{d{#1}}{d{#2}}}

\newcommand{\fracp}[2]{\frac{\partial{#1}}{\partial{#2}}}

\newcommand{\rlr}[1]{#1} 
\newcommand{\rla}[1]{#1} 

\makeatletter
\newenvironment{newfigure}
  {\renewcommand{\fnum@figure}{\textcolor{blue}{\bfseries Figure~\thefigure}}%
   \begin{figure}}
  {\end{figure}}
\newenvironment{newfigure*}
  {\renewcommand{\fnum@figure}{\textcolor{blue}{\bfseries Figure~\thefigure}}%
   \begin{figure*}}
  {\end{figure*}}
\newenvironment{revfigure}
  {\renewcommand{\fnum@figure}{\textcolor{blue}{\bfseries Figure~\thefigure}}%
   \begin{figure}}
  {\end{figure}}
\newenvironment{revfigure*}
  {\renewcommand{\fnum@figure}{\textcolor{blue}{\bfseries Figure~\thefigure}}%
   \begin{figure*}}
  {\end{figure*}}
\makeatother

\received{--}
\revised{--}
\accepted{--}
\submitjournal{ApJ}

\shorttitle{In-Situ Formation of the CCKB}
\shortauthors{Li and Chiang}

\begin{document}

\title{In-Situ Formation of the Cold Classical Kuiper Belt}

\correspondingauthor{Rixin Li}
\email{rixin@berkeley.edu}
\author[0000-0001-9222-4367]{Rixin Li
\begin{CJK*}{UTF8}{gkai}(李日新)\end{CJK*}
}
\altaffiliation{51 Pegasi b Fellow}
\affiliation{Department of Astronomy, Theoretical Astrophysics Center, and Center for Integrative Planetary Science, University of California Berkeley, Berkeley, CA 94720-3411, USA}

\author[0000-0002-6246-2310]{Eugene Chiang
\begin{CJK*}{UTF8}{bkai}(蔣詒曾)\end{CJK*}}
\affiliation{Department of Astronomy, Theoretical Astrophysics Center, and Center for Integrative Planetary Science, University of California Berkeley, Berkeley, CA 94720-3411, USA}
\affiliation{Department of Earth and Planetary Science, University of California Berkeley, Berkeley, CA 94720-4767, USA}

\begin{abstract}
Cold Classical Kuiper belt objects (CCKBOs) are considered first-generation planetesimals that formed 42--47 au from the Sun and remained untouched since. Formation is thought to proceed by clumping of dust particles in protoplanetary disk gas by the streaming instability, followed by gravitational collapse. Previous calculations along these lines are inconsistent with the CCKB's supposedly pristine nature, because they assume orders of magnitude more solid mass than is actually present in the CCKB (a few thousandths of an Earth mass) and do not explain how to expel the $> 99$\% extra mass. Here we show from 3D numerical simulations of dust and gas that the total mass in CCKBOs, their characteristic sizes of $\sim$100 km, and the \rla{relative proportion of prograde to retrograde binaries} can all be reproduced at the tail end of the solar nebula's life, when it contained just 2--5\% of its original (minimum-mass) gas. As a solar metallicity's worth of mm-sized solids drains out from 42--47 au from nebular headwinds, about 1\% of the dust collapses into planetesimals that remain behind in the CCKB region. Binarity is guaranteed from a simple analytic estimate, confirmed numerically, of the spin angular momentum in clumps seeded by the streaming instability. We show that other formation scenarios, including trapping of dust within a gas pressure bump, fail to reproduce the low-mass CCKB. Outstanding problems are identified.
\end{abstract}

\section{Introduction}
\label{sec:intro}

Of the various components of the Kuiper belt (for reviews, see \citealt{Morbidelli2020}; \citealt{Gladman_Volk2021}), Cold Classical Kuiper belt objects (CCKBOs) are thought to be the most dynamically pristine. Confined to a narrow range of heliocentric semimajor axes ($a \approx 42$--47 au), and having eccentricities and inclinations of less than a few percent (e.g.~\rla{\citealt{Brown2001}}; \citealt{Van_Laerhoven2019}; \rla{\citealt{Batygin2020};} \citealt{Kavelaars2021}; \citealt{Huang2022}; \citealt{Fraser2024}), the CCKB appears to have avoided the upheavals that created other denizens of the outer solar system (resonant and scattered Kuiper belt objects, and the inner and outer Oort clouds). Underscoring the CCKB's unique history and dynamical isolation are its distinct surface colors (e.g.~\citealt{Gulbis2006}, and references therein; \citealt{Fraser2017}; \citealt{Fraser2021}) and its high incidence of wide-separation, equal-size binaries (e.g.~\citealt{Noll2020}) whose survival depends on avoiding collisional impacts, and planetary encounters that transported other, non-CC KBOs across trans-Neptunian space (e.g.~\citealt{Nesvorny2019b}; \citealt{Nesvorny2022}; and references therein). For these reasons, it is suspected the CCKB formed in situ, and remained relatively untouched since.

The streaming instability (SI; \citealt{Youdin_Goodman_2005}) has emerged as a leading candidate for how CCKBOs and planetesimals in general form. The SI is a linear instability in protoplanetary disks that concentrates solid particles aerodynamically dragged by gas, thereby facilitating their gravitational collapse into planetesimals \citep[e.g.][]{Johansen2007a, Simon2017, LY21}. \citet{Nesvorny2019} demonstrated that the SI generally produces binary planetesimals whose properties resemble those of CCKB binaries.  About 80\%/20\% of CCKB binaries are prograde/retrograde \citep{Grundy2019}; this proportion can be traced to the vorticity distribution in 3D streaming turbulence. The SI further yields a mass spectrum of planetesimals that is top-heavy and cuts off exponentially at masses corresponding to solid bodies $\sim$100 km in diameter (\citealt{Schafer2017, Li2019, Klahr_Schreiber_2020, Gerbig2023}). This theoretical mass function (a.k.a.~size-frequency distribution) is similar to the observed mass functions in the CCKB and other portions of the Kuiper belt (\citealt{Kavelaars2021,Petit2023,Napier2023}) and asteroid belt (\citealt{Morbidelli2009, Morbidelli2020}).

All told, the CCKB presents an excellent (and arguably currently our only) means of empirically testing theories for the formation of the first-generation planetesimals. We aim here to assess whether and how the SI can reproduce the CCKB's remarkably low mass.  Today, the CCKB has a total mass of $\sim$$0.0021 M_\oplus$, uncertain by a factor of 3 to either side \citep{Petit2023,Napier2023}. This present-day mass is probably not much lower than the mass at the time of the CCKB's formation, insofar as CCKBOs are collisionless and relatively immune to disturbances by the giant planets. In the planetary migration scenarios considered by \citet{Nesvorny2015}, the CCKB loses up to half of its mass from dynamical erosion over 4 Gyr.

We will adopt a primordial CCKB mass of $M_{\rm CCKB} = 0.003 M_\oplus \pm 0.5$ dex; this is the mass contained in $\sim$100-km sized planetesimals within a heliocentric annulus of mean radius $r = 45$ au and radial width $\Delta r = 5$ au.  The implied disk surface density in solids is three orders of magnitude smaller than the corresponding solid surface density in typical solar nebula models.  The SI-based binary formation simulations of \citet{Nesvorny2019} assumed solid surface densities comparable to the latter, thereby forming a belt much too massive compared to the actual CCKB \rla{(see also \citealt{Simon2024})}. \citet{Gomes2021} also noted this shortcoming, which motivated their proposal that the CCKB did not form in situ, but was transported to its current location as part of the migration history of the giant planets. Our paper revisits the idea of in-situ formation to ask what protoplanetary disk conditions are needed for the SI to create the CCKB. The low mass and dynamically cold nature of the CCKB suggest it formed late, as the last of the solar nebula was about to dissipate, and perhaps also after the chaos that generated the dynamically hotter components of the Kuiper belt. Formation at a disk age older than a few Myr is consistent with the low bulk densities of CCKBOs, which must have avoided internal heating and melting from live $^{26}$Al \citep{Bierson2019}. 

The ring occupied by CCKBOs calls to mind the dust rings observed in young protoplanetary disks by ALMA (Atacama Large Millimeter Array; e.g. \citealt{Andrews2018, Huang_DSHARP_2018}). The ALMA rings coincide with local gas pressure ``bumps'' --- maxima in the radial pressure profile, of unknown origin --- which collect dust particles, possibly millimeters in size, by aerodynamic drag (\citealt{Stadler2025}, and references therein).  \rla{The CCKB, which contains far less solid mass than in a typical ALMA ring, might have formed as a scaled-down version of the latter, with planetesimals coagulating from dust concentrated within a gas pressure maximum.} We began the present study by considering this hypothesis, but ultimately abandoned it. Using simulations similar to those of \citet{LY21}, and assuming no intrinsic turbulence in the gas, we found that dust particles drifting into a pressure bump agglomerated into a single giant planetesimal --- whereas the CCKB contains on the order of thousands of 100-km planetesimals. Streaming turbulence, which serves to keep particles interspersed in gas, could not be sustained within the pressure maximum, which by construction weakens the radial pressure gradient needed to drive the SI.  \citet{Lee2022} found the same outcome of excessively efficient coagulation in laminar pressure bumps. Including a separate source of gas turbulence in the pressure bump only raises the solid mass needed for gravitational collapse  and yields a planetesimal belt much more massive than the CCKB, as we will quantify later in this paper.

For the bulk of our study we investigate instead the simpler scenario of dust particles drifting radially inward through a smooth (no bump) disk, clumping into proto-planetesimals from the SI as they pass through the CCKB region of $\sim$42--47 au. We will successfully reproduce several observed properties of the CCKB, including its low solid surface density \rla{and binary properties}. However, because planetesimals in this steady drift scenario form over an arbitrarily large range of heliocentric distances, the question of why the belt truncates at its outer edge of $\sim$47 au is unanswered (on the other hand, the inner edge at $\sim$42 au is adequately explained by sculpting from various giant planet secular resonances; \citealt{Knezevic1991, Holman_Wisdon_1993, Duncan1995}). Also left unspecified is how our modeled radial influx of low-density particles fits with the overall timeline of events in the outer solar system --- why, e.g., prior episodes of particle drift did not also form planetesimals to create a more massive and/or extended belt.  Our main research product is limited to a determination of the surface densities in disk gas and solids needed for the SI to reproduce the CCKB's surface density today.

This paper is organized as follows.  Section \ref{sec:method} describes how we set up our smooth disk (no bump) simulations. Section \ref{sec:results} reports on the numbers, masses, and potential binarity, including obliquities, of planetesimals formed by the SI acting on particles drifting inward through smooth disks. We include in that section an analytic, order-of-magnitude derivation of the spin angular momentum of SI-formed planetesimals, explaining why the SI naturally creates binaries. Section \ref{sec:pure_GI} considers alternative formation scenarios for the CCKB and  presents an extra set of simulations that explore particle concentration and clumping within turbulent gas pressure bumps. We summarize in Section \ref{sec:summary}.

\section{Method for Smooth Disk (No Bump) Simulations}
\label{sec:method}

To simulate gas and solid particles with mutual drag and self-gravity, we use the \texttt{ATHENA} code \citep{Stone2008} outfitted with a particle module by \citet{Bai2010} and a particle self-gravity extension by \citet{Simon2016}.  Our methods are similar to those of \citet{Li2019}, but are customized for a late-stage, low-mass disk. We simulate how mm-sized particles drift via gas drag through a test volume modeled after the Cold Classical Kuiper belt, and assess how these particles can clump en route
into much larger CCKBOs, thereby nearly halting their drift and remaining within the belt volume. The gas is assumed to have a global radial pressure profile that decreases monotonically outward, and would be laminar were it not for stirring from the solids (by contrast, our simulations in Section \ref{subsec:bump_sims} model a radial pressure bump in gas which is intrinsically turbulent).

Section \ref{subsec:gas-dust} summarizes the equations solved and our numerical boundary conditions.  Section \ref{subsec:setup} details physical input parameters and code units. Section \ref{subsec:setup2} lists simulation domain sizes, the number of dust super-particles used, and run times. \rla{Note that this section \ref{sec:method} describes how we construct simulations conducted at our standard spatial resolution. Supplemental runs at higher spatial resolution, initialized using the results of a standard run, are detailed in Section \ref{subsec:zoom_sim}.}

\begingroup 
\setlength{\medmuskip}{0mu} 
\begin{deluxetable*}{cccccccccc}[ht]
  \tablecaption{Smooth Disk (No Bump) Simulation Parameters}\label{tab:paras}
  \tablecolumns{10}
  \tablehead{
    \colhead{(1)} &
    \colhead{(2)} &
    \colhead{(3)} &
    \colhead{(4)} &
    \colhead{(5)} &
    \colhead{(6)} &
    \colhead{(7)} &
    \colhead{(8)} &
    \colhead{(9)} &
    \colhead{(10)} \\
    \colhead{\texttt{Run}} &
    \colhead{$\mathcal{F}_{\rm gas}$} &
    \colhead{$Z$} &
    \colhead{$Q$} &
    \colhead{$\Pi$} &
    \colhead{$\uptau_{\rm s,1mm}$} &
    \colhead{$t_{\rm sed}$} &
    \colhead{$M_{\rm init,belt}$} &
    \colhead{$M_{\rm final,belt}$} &
    \colhead{$R_{\rm plan}$} \\
    \colhead{} &
    \colhead{} &
    \colhead{} &
    \colhead{} &
    \colhead{} &
    \colhead{} &
    \colhead{[$\Omega_0^{-1}$]} &
    \colhead{[$M_{\oplus}$]} &
    \colhead{[$10^{-3}M_{\oplus}$]} &
    \colhead{[km]}
  }
  \startdata
  \hline
  \texttt{A}       & $1\%$ & $0.01$   & $1252.8$ & $0.056$ & $2.16$ & $35$ & $0.0386$ & $0$             & …       \\
  \texttt{B}       & $2\%$ & $0.01$   & $626.4$  & $0.056$ & $1.08$ & $35$ & $0.0772$ & $0$             & …       \\
  \texttt{C}       & $3\%$ & $0.01$   & $417.6$  & $0.056$ & $0.72$ & $35$ & $0.116$  & ($1.32$)        & $(114)$ \\
  \texttt{D}       & $5\%$ & $0.01$   & $250.6$  & $0.056$ & $0.43$ & $40$ & $0.193$  & $1.43$ ($5.28$) & $117$   \\
  \texttt{\rla{D-Zoom-P}} & \rla{$5\%$} & \rla{$0.01$} & \rla{$250.6$} & \rla{$0.056$} & \rla{$0.43$} & \nodata & \nodata & \rla{$11.8$} & \rla{$132$} \\
  \texttt{\rla{D-Zoom-O}} & \rla{$5\%$} & \rla{$0.01$} & \rla{$250.6$} & \rla{$0.056$} & \rla{$0.43$} & \nodata & \nodata & \rla{$8.9$} & \rla{$118$} \\
  \hline\hline
  \texttt{A-hiZ}   & $1\%$ & $0.05$   & $1252.8$ & $0.056$ & $2.16$ & $35$ & $0.386$  & $0$             & …  \\
  \texttt{B-hiZ}   & $2\%$ & $0.025$  & $626.4$  & $0.056$ & $1.08$ & $35$ & $0.193$  & ($3.86$)        & (99)  \\
  \texttt{C-hiZ}   & $3\%$ & $0.0167$ & $417.6$  & $0.056$ & $0.72$ & $35$ & $0.193$  & $5.39$ ($8.10$) & 172  \\
  \enddata
  \tablecomments{Columns: 
  (1) run name;
  (2) ratio of gas surface density $\Sigma_{\rm g}$ relative to MMSN at $r = 45$ au ($\Sigma_{\rm g,MMSN} = 7.3$ g/cm$^2$); 
  (3) height-integrated metallicity $\Sigma_{\rm p}/\Sigma_{\rm g}$; 
  (4) Toomre $Q$ for the gas disk;
  (5) dimensionless radial pressure gradient;
  (6) dimensionless stopping time for dust particles of size $a = 1$ mm;
  (7) duration of the initial transient sedimentation phase, interpolated from Table 2 in \citet{LY21};
  (8) initial ``belt''  mass in particles, computed by scaling our shearing box to an annulus of radial width $5$ au centered at $r = 45$ au;
  (9) final ``belt'' mass (scaled as in (8)) in persistent clumps formed from the gravitational collapse of particles, with parenthetical value accounting for all clumps both persistent and transient; 
  (10) planetesimal radius computed from the most massive persistent clump, assuming the clump forms an equal-mass binary, evaluated at the end of the run. The listed radius is of one binary component, of mean density $\rho_\bullet = 1$ g/cc. Parenthetical value is derived from the most massive, transient clump, evaluated at its peak mass before it dissipates. 
  Simulation videos are available at \href{https://www.rixinli.com/cckb}{rixinli.com/CCKB}.
  }
\end{deluxetable*}
\endgroup

\subsection{\texorpdfstring{Governing Equations and Boundary Conditions}{Governing Equations and Boundary Conditions}}  
\label{subsec:gas-dust}

We simulate a vertically stratified volume of the protoplanetary disk using a local shearing box \citep{Hawley1995}.  The simulation domain is centered on the disk midplane at a fiducial stellocentric radius $r$ where the local orbital frequency is $\Omega_0$. 
Positions relative to the box center are described by local Cartesian coordinates $\{x, y, z \}$ in the radial, azimuthal, and vertical directions, respectively.  In this domain, \texttt{ATHENA} solves the equations of gas dynamics
\begin{align}
  \fracp{\rho_{\rm g}}{t} + \nabla \cdot (\rho_{\rm g} \bm{u}) &= 0 \label{eq:gascon}\\
  \begin{split}\label{eq:gasmom}
    \fracp{(\rho_{\rm g} \bm{u})}{t} + \nabla\cdot(\rho_{\rm g} \bm{u}\bm{u} + P\bm{I}) &=\\
    \rho_{\rm g} \biggl[ 2\bm{u}\times\bm{\Omega}_0 + 3{\Omega}_0^2 \bm{x} - {\Omega}_0^2 \bm{z} &- \nabla\Phi_{\rm sg} \biggr] + \rho_{\rm p} \frac{\bar{\bm{v}} - \bm{u}}{t_\mm{stop}}
  \end{split} \\
  P &= \rho_{\rm g} c_{\rm s}^2 \,,
  \end{align}
in addition to the equation of motion for each solid super-particle (indexed by $i$)
\begin{align}
  \begin{split}\label{eq:ithpar}
    \fracd{\bm{v}_i}{t} = 2\bm{v}_i\times\bm{\Omega}_0+3{\Omega}_0^2 \bm{x}_i &- {\Omega}_0^2 \bm{z}_i \\
    -\frac{\bm{v}_i - \bm{u}}{t_{\mm{stop},i}} &- \nabla\Phi_{\rm sg} - 2\eta v_{\rm K} \Omega_0 \hat{x} \,,
  \end{split} 
  \end{align}
and Poisson's equation for super-particle self-gravity
  \begin{align}
  \nabla^2 \Phi_{\rm sg} &= 4 \pi G \rho_{\rm p} \,.
\end{align}
Here $\rho_{\rm g}$, $\bm{u}$, $P$, and $c_{\rm s}$ are the density, velocity, pressure, and sound speed of the isothermal un-magnetized gas, $\bm{I}$ is the identity matrix, $\bm{\Omega}_0 = \Omega_0 \hat{z}$, $\rho_{\rm p}$ and $\bar{\bm{v}}$ are the volumetric density and average velocity of particles in a grid cell, 
$t_{\rm stop}$ is the dimensional stopping time of particles in gas (assumed equal for all particles),  $\Phi_{\rm sg}$ is the self-gravitational potential of particles, and $\eta$ parameterizes the strength of the background radial pressure gradient which determines how much slower gas orbits relative to the Keplerian velocity $v_{\rm K}$ (see \citealt{Bai2010} for more details). 

For the gas, we employ shearing-periodic boundary conditions (BCs) in the radial direction, and periodic BCs in the azimuthal direction.  Outflow BCs are imposed in the vertical direction, with the gas density extrapolated into ghost zones to maintain hydrostatic balance, and gas inflow prohibited \citep{Simon2011, Li2018}. For particles, periodic BCs are adopted in azimuth, and outflow BCs are imposed vertically (particles that exit the box vertically do not return). A radial shearing-periodic BC for particles is used only initially when waiting for transients in the particle field to damp away; afterward, for the remainder of the simulation, we switch to using an outflow BC for particles such that particles that exit the box radially do not return. More details on the initialization procedure are given below in the subsections below.

\subsection{Dimensionless Parameters, Initial Conditions, and Code Units}
\label{subsec:setup}
Four dimensionless parameters characterize the simulations: the non-dimensional particle stopping time or Stokes number
\begin{equation}
  \uptau_{\rm s} = \Omega_0 t_{\rm stop} \,;
\end{equation}
the ratio of the average, vertically integrated surface density of solids to that of the gas, a.k.a.~the metallicity
\begin{equation}
  Z = \frac{\Sigma_{\rm p}}{\Sigma_{\rm g}} \,;
\end{equation}
the global radial pressure gradient 
\begin{equation}
  \Pi \equiv \frac{\eta \Omega_0 r}{c_{\rm s}} \equiv -\frac{1}{2}\frac{c_{\rm s}}{\Omega_0 r}\fracp{\,\ln{P}}{\,\ln{r}} \,;
\end{equation}
and the initial strength of gas self-gravity \citep{Simon2016}
\begin{equation}
  \widetilde{G} \equiv \frac{4 \pi G \rho_{\rm g,0}}{\Omega_0^2} \label{eq:tildeG}
\end{equation}
where $G$ is the gravitational constant and $\rho_{\rm g,0}$ is the initial gas density at the midplane.  Because gas motions are highly subsonic, gas densities hardly change in the simulation and the midplane gas density is effectively fixed at $\rho_{\rm g,0}$. 
An equivalent parameter to $\widetilde{G}$ is Toomre's
\begin{equation} \label{eq:toomre}
  Q = \frac{c_{\rm s}\Omega_0}{\pi G \Sigma_{\rm g}} = \sqrt{\frac{8}{\pi}} \,\widetilde{G}^{-1}\,.
\end{equation}
Note that $\widetilde{G}\leftrightarrow Q$ describes the self-gravity of the background gas, not the particles.

We draw parameters appropriate to $r = 45$ au, the approximate mean radius of the Cold Classical Kuiper belt.  The minimum-mass solar nebula (MMSN) provides a reference gas surface density $\Sigma_{\rm g,MMSN} = 2200 \,(r/\text{au})^{-3/2}$ g cm$^{-2} = 7.3$ g cm$^{-2}$, of which only a fraction may be present in a dissipating, late-stage, low-mass disk: $\Sigma_{\rm g} = \mathcal{F}_{\rm gas} \Sigma_{\rm g,MMSN}$. We vary $\mathcal{F}_{\rm gas}$ from $1\%$ to $5\%$. The local disk temperature is $T = 120 \,( {r}/{\text{au}})^{-3/7} ~\text{K} = 23.5 ~\text{K}$ \citep{Chiang2010}, implying a gas disk aspect ratio of $h = (c_{\rm s}/\Omega_0)/r = H/r = 0.065$.  Given the above parameter choices, and assuming a gas surface density profile that is locally flat at $r = 45$ au, we compute the dimensionless pressure gradient $\Pi = 0.056$, fixed for all our smooth disk simulations; see Table \ref{tab:paras} for this and other parameters. Although we account for a radial temperature gradient when evaluating $\Pi$, our shearing box simulations use a strictly isothermal gas.

Initial gas densities $\rho_{\rm g}$ are uniform in $x$ and $y$, and decrease in $z$ away from the midplane as a Gaussian with scale height $H$.  The above parameters define the midplane gas density $\rho_{\rm g,0}$ and thus $Q \leftrightarrow \widetilde{G}$.  In units of $\rho_{\rm g,0}$, the Roche density (above which a material becomes sufficiently self-gravitating to resist stellar tidal shear) is
\begin{equation} \label{eq:roche}
\frac{\rho_{\rm R}}{\rho_{\rm g,0}} = \frac{9\Omega_0^2}{ 4\pi G \rho_{\rm g,0}} = \frac{9 \sqrt{2\pi} }{4} Q = 9\,{\widetilde{G}}^{-1} \,.
\end{equation}

Dust particles are assumed identical with dimensionless stopping times 
\begin{equation}
  \uptau_{\rm s} = \frac{\pi}{2} \frac{\rho_\bullet a}{\Sigma_{\rm g}} 
\end{equation}
evaluated in the Epstein regime for material density $\rho_\bullet = 1$ g cm$^{-3}$ and fixed particle size $a = 1$ mm. 
Initial particle densities are uniform in $x$ and $y$, and follow a Gaussian 
\begin{equation}
  \rho_{\rm p} = \frac{\Sigma_{\rm p}}{\sqrt{2\pi}H_{\rm p}}\exp{\left(\frac{-z^2}{2H_{\rm p}^2} \right)} 
\end{equation}
in $z$ with $H_{\rm p} = 0.025H$ and $\Sigma_{\rm p} = Z \Sigma_{\rm g}$ (see Table \ref{tab:paras} for $Z$).  The velocities of gas and particles are initialized with equilibrium drift velocities  accounting for the momentum backreaction of particles on gas \citep{Nakagawa1986}. Note that these initial conditions will be modified in a pre-run procedure described in \S\ref{subsec:setup2}.

Code units include the local orbital timescale $\Omega_0^{-1} = 1$, the gas sound speed $c_{\rm s} = 1$, the gas scale height $H = c_{\rm s}/\Omega_0 = 1$, and the initial midplane gas density $\rho_{\rm g,0} = 1$.  

\subsection{Box Sizes, Particle Numbers, and Run Times}
\label{subsec:setup2}

The radial, azimuthal, and vertical extents of our simulation domain are $L_X \times L_Y \times L_Z = (1.7 \times 0.15 \times 0.2) H^3$.  The radial extent $L_X$ is chosen to match the width ($\Delta r \simeq 5$ au) of today's Classical Kuiper belt.  Grid cells are $\Delta x = H/1040$ on a side.  Our vertical outflow BCs have been shown to minimize boundary artifacts and produce convergent results with different box heights \citep{Li2018}. 

The number of Lagrangian dust super-particles is 
\begin{equation}\label{eq:N_p}
  N_{\rm p} = n_{\rm p}\cdot \frac{L_X \cdot L_Y \cdot 2H_{\rm p}}{(\Delta x)^3} \approx 5.74\times 10^{7} 
\end{equation}
where $n_{\rm p}=4$ (see also Eq.~5 of \citealt{Li2018}).  Unlike in some other studies, dust self-gravity is continuously accounted for in our simulations (i.e.~it is not suddenly switched on at some intermediate time during the run; see \citealt{Gerbig2023} for a discussion of this point).  If gravitational collapse occurs, we use the clump-finding tool \texttt{PLAN} \citep{Li2019,PLAN,Gole2020} to identify and track self-bound clumps.

Our simulations are first run for $t_{\rm sed} = 35$--$40\Omega_0^{-1}$ (Table \ref{tab:paras}) using radial shearing-periodic boundary conditions for the particles. During this transient sedimentation phase, particles settle vertically into a quasi-equilibrium regulated by streaming instability turbulence (see Fig.~3 and Table 2 in \citealt{LY21}). After transients damp away, we switch to outflow boundary conditions for the particles, and run for as long as it takes all the particles to drift radially inward and exit the simulation domain, about $150\Omega_0^{-1}$.

To place our simulations into context with the Cold Classical Kuiper belt, we define an initial ``belt'' mass
\begin{equation}
M_{\rm init,belt} \equiv 2\pi \Sigma_{\rm p} r \Delta r
\end{equation}
where $r = 45$ au and $\Delta r = 5$ au. This mass accounts for the full $2\pi$ azimuthal extent of the belt, as distinct from the simulation box which spans only $L_Y/r = 0.15H/r = 0.00975$ rad (and is actually Cartesian, the curvilinear terms having been suppressed in the shearing-box equations of motion). Planetesimals that form within the simulation box are used to evaluate a final belt mass $M_{\rm final,belt} = M_{\rm init,belt} \times M_{\rm final}/M_{\rm init}$, where $M_{\rm init}$ and $M_{\rm final}$ are the initial and final solid masses in the box (initially in particles and finally in planetesimals, the majority of particles having drifted out of the box).

\begin{figure}
  \centering
  \includegraphics[width=\linewidth]{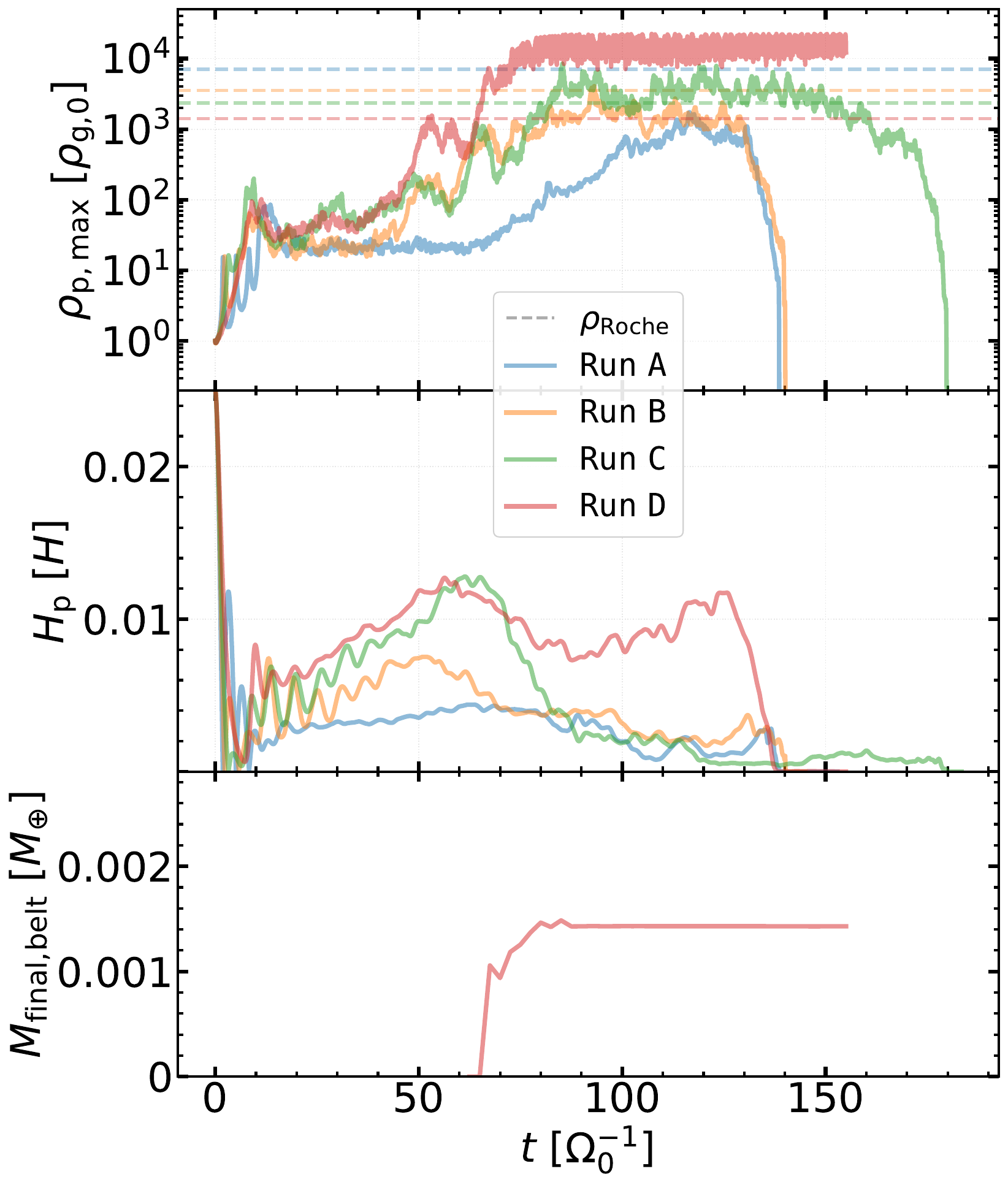}
  \caption{\textit{Top panel:} Maximum particle densities vs.~time for \texttt{Runs A-D} which progressively increase $M_{\rm init,\!belt}$ (see Table \ref{tab:paras}). At fixed $Z$ and initial $H_{\rm p} = 0.025 H$, all runs start with the same $\rho_{\rm p,max} / \rho_{\rm g,0} = 0.4$. For $t \leq t_{\rm sed} = 35$--40 $\Omega_0^{-1}$, radial shearing periodic boundary conditions are implemented and the particle layer is allowed to vertically settle into a quasi-equilibrium; afterward, radial outflow boundary conditions are implemented and the simulation is run until all the particles (but not necessarily the planetesimals) drift out of the simulation box.  Dashed horizontal lines mark Roche densities $\rho_{\rm R}$, color-coded by run.  All \texttt{Runs A-D} exhibit particle clumping due to the SI, but only in \texttt{Runs C-D} do overdensities exceed the Roche density and gravitationally collapse, either into transient clumps ultimately disrupted by tidal shear (\texttt{Runs C-D}), or clumps that survive the run duration (\texttt{Run D} only).  \textit{Middle panel:} Particle scale heights $H_{\rm p}$, evaluated as the standard deviation (square root of the variance) of vertical particle positions. Oscillations reflect particles vertically settling, overshooting the midplane, and re-settling.  \textit{Bottom panel:} The mass in bound, surviving clumps formed within the simulation box of azimuthal length $L_Y$, scaled up by $2\pi r / L_Y = 2\pi / 0.00975 = 644$ to estimate the planetesimal mass within an annulus having the full circumference $2\pi r$. Only \texttt{Run D} produces a surviving clump, and a final belt mass of $M_{\rm final,belt} = 0.0013 M_\oplus$ that falls within the range of estimated present-day CCKB masses ($0.0021 M_\oplus \pm 0.5$ dex).
  \label{fig:par_stats}}
\end{figure}

\begin{figure*}
  \centering
  \includegraphics[width=0.925\linewidth]{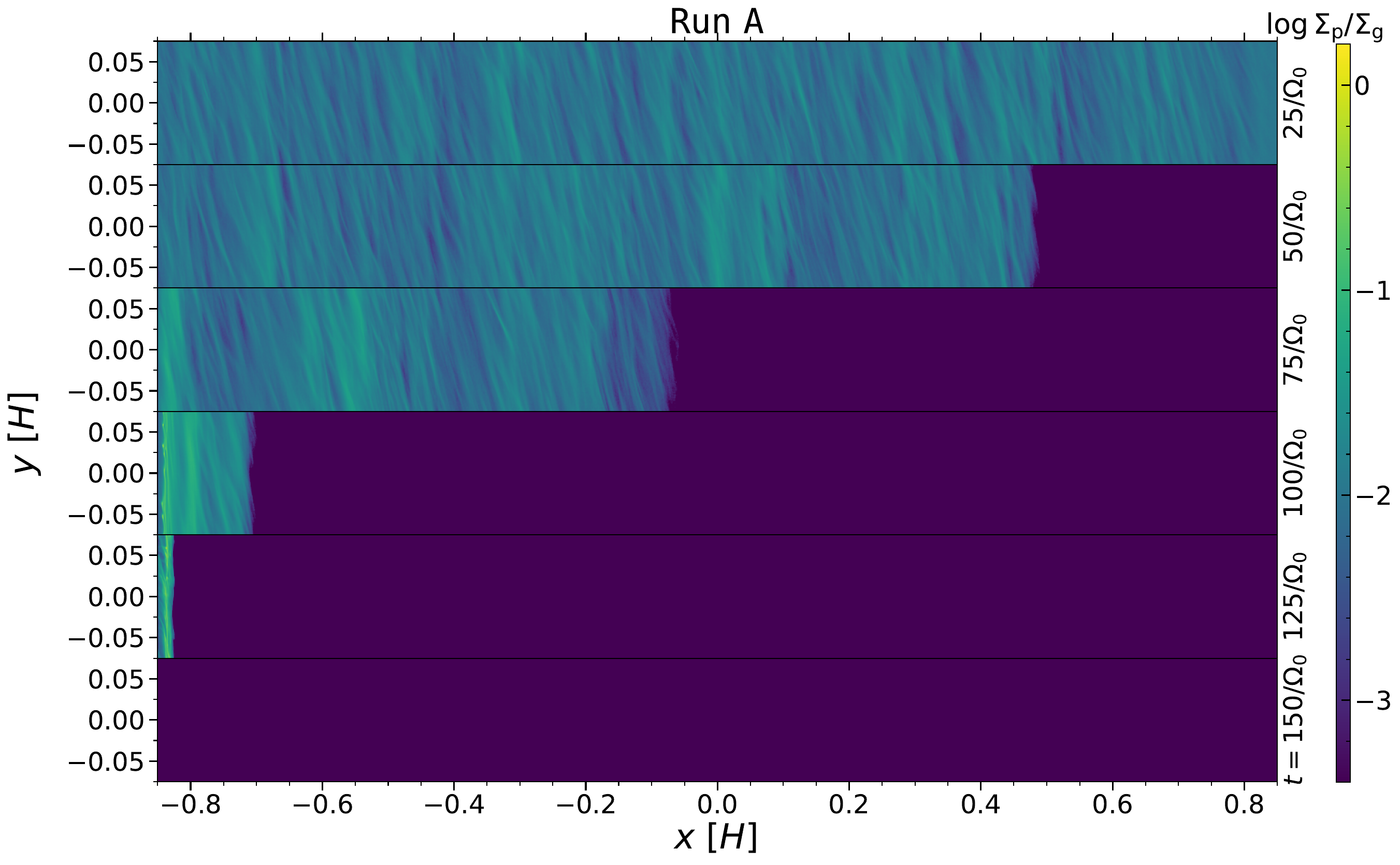}
  \includegraphics[width=0.925\linewidth]{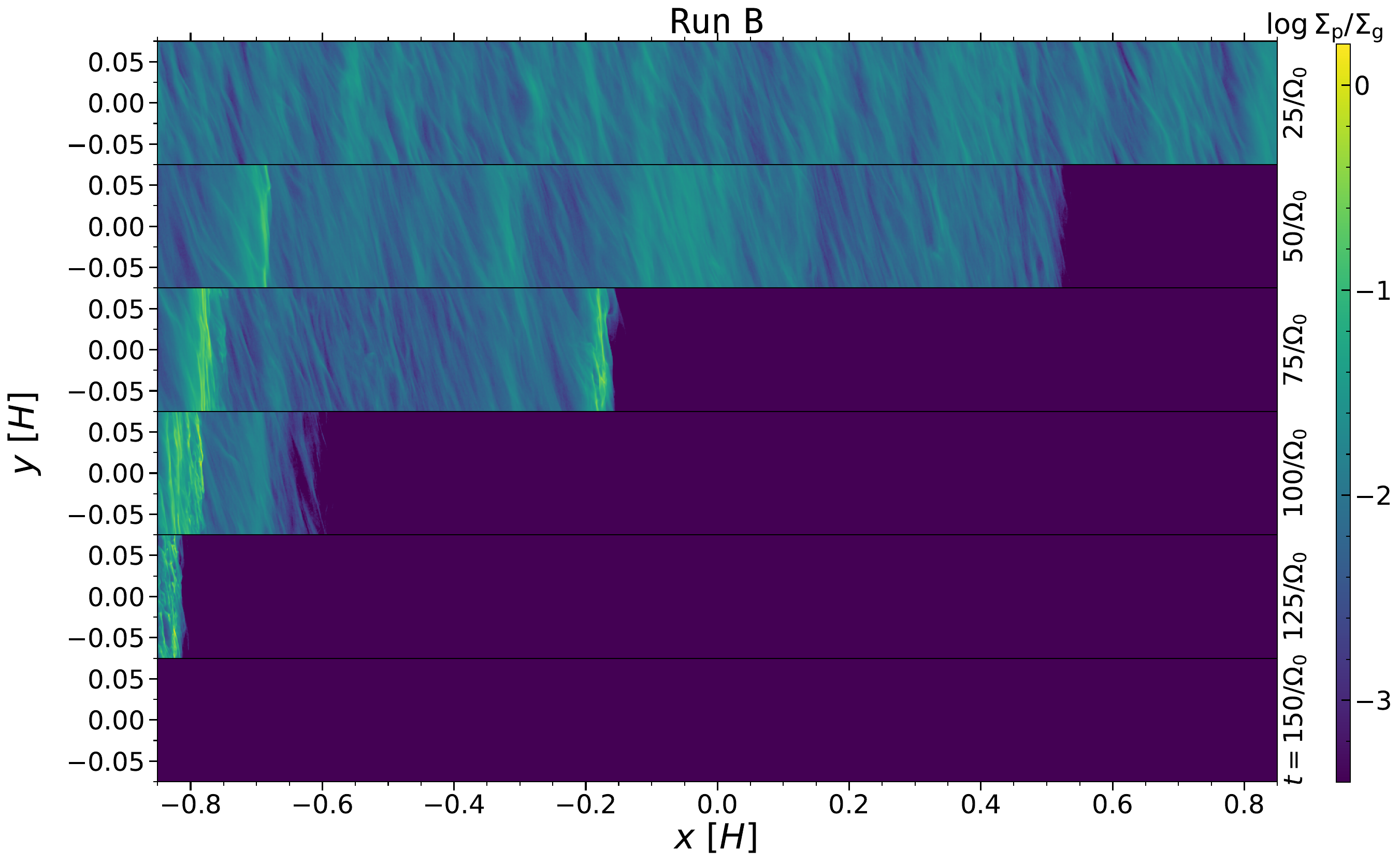}
  \caption{Snapshots of particle surface density $\Sigma_{\rm p}$, normalized to the gas surface density $\Sigma_{\rm g}$ which is nearly constant because the gas behaves approximately incompressibly, for \texttt{Run A} (top) and \texttt{Run B} (bottom). Snapshot times are listed vertically along the right-hand axis. The simulations are run until all particles drift out of the box (toward the left). For these two runs, filaments form but do not spawn planetesimals.  Simulation videos are available at \href{https://www.rixinli.com/cckb}{rixinli.com/CCKB}.
  \label{fig:new_snapshots1}}
\end{figure*}

\begin{figure*}
  \centering
  \includegraphics[width=0.925\linewidth]{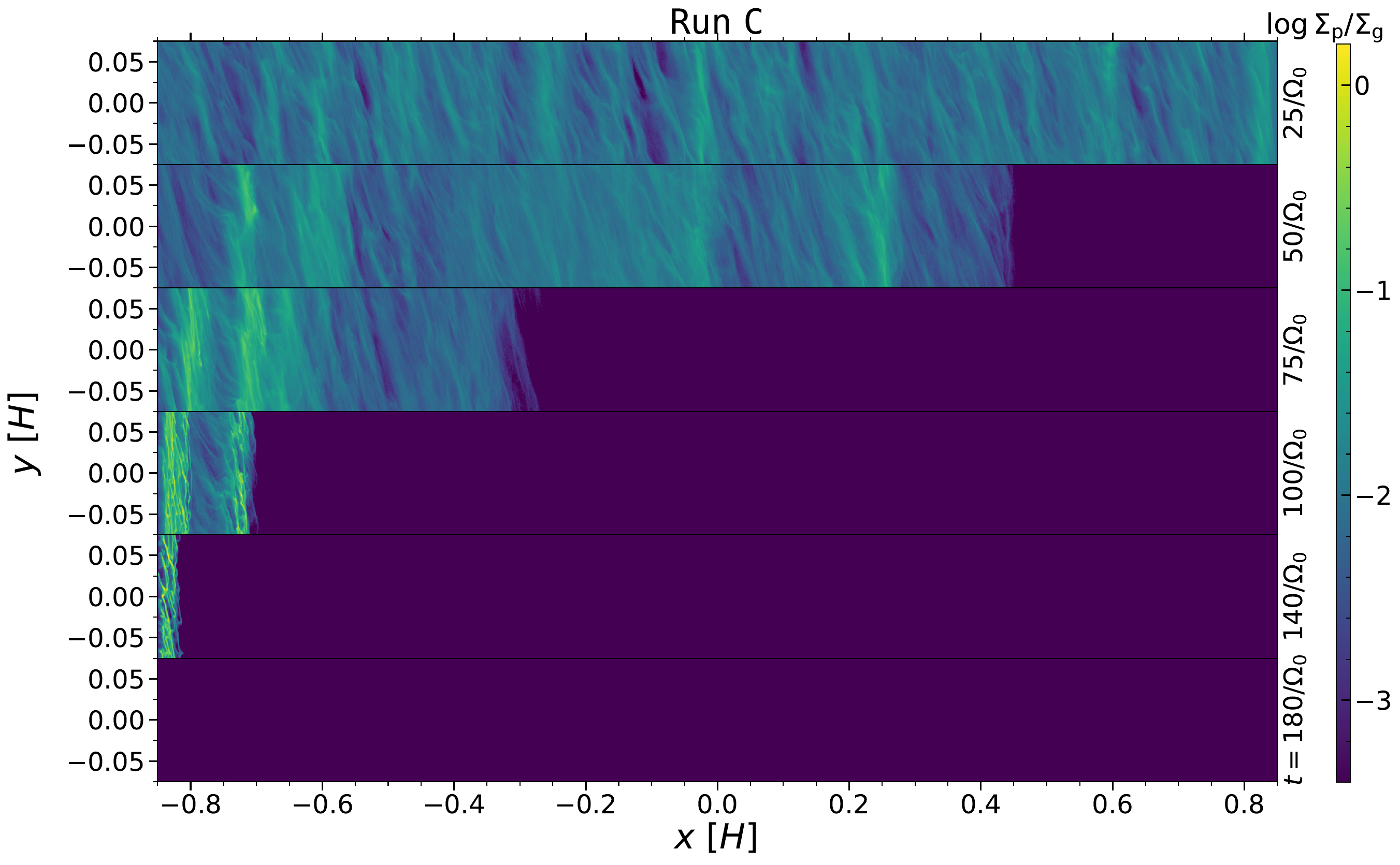}
  \includegraphics[width=0.925\linewidth]{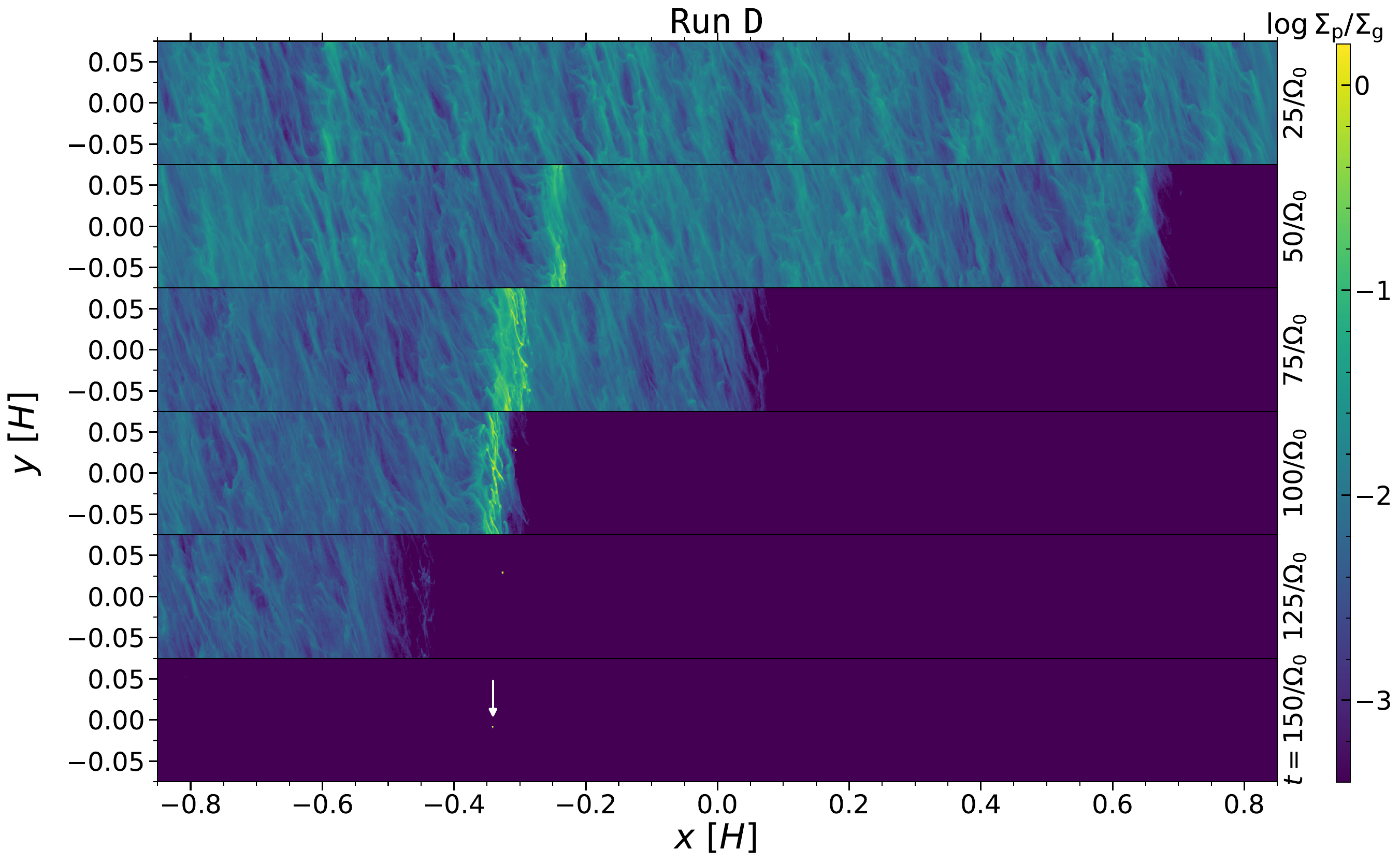}
  \caption{Same as Fig.~\ref{fig:new_snapshots1} but showing snapshots from \texttt{Run C} (top) and \texttt{Run D} (bottom). In both runs, clumps form out of filaments but ultimately disperse from tidal shear. In \texttt{Run D}, one clump is massive and dense enough to survive, and remains within the simulation domain --- see the bright point at $x (H_{\rm g}) \approx -0.3$ to $-0.35$ --- after all other solids drift away. The properties of the surviving clump are consistent with those of the CCKB. We use the transient clumps, which we argue may actually survive in a higher-resolution simulation, to estimate an upper limit on the planetesimal belt mass (listed in parentheses under $M_{\rm final,belt}$ in Table \ref{tab:paras}).  Simulation videos are available at \href{https://www.rixinli.com/cckb}{rixinli.com/CCKB}.
  \label{fig:new_snapshots2}}
\end{figure*}

\begin{figure}
  \centering
  \includegraphics[width=\linewidth]{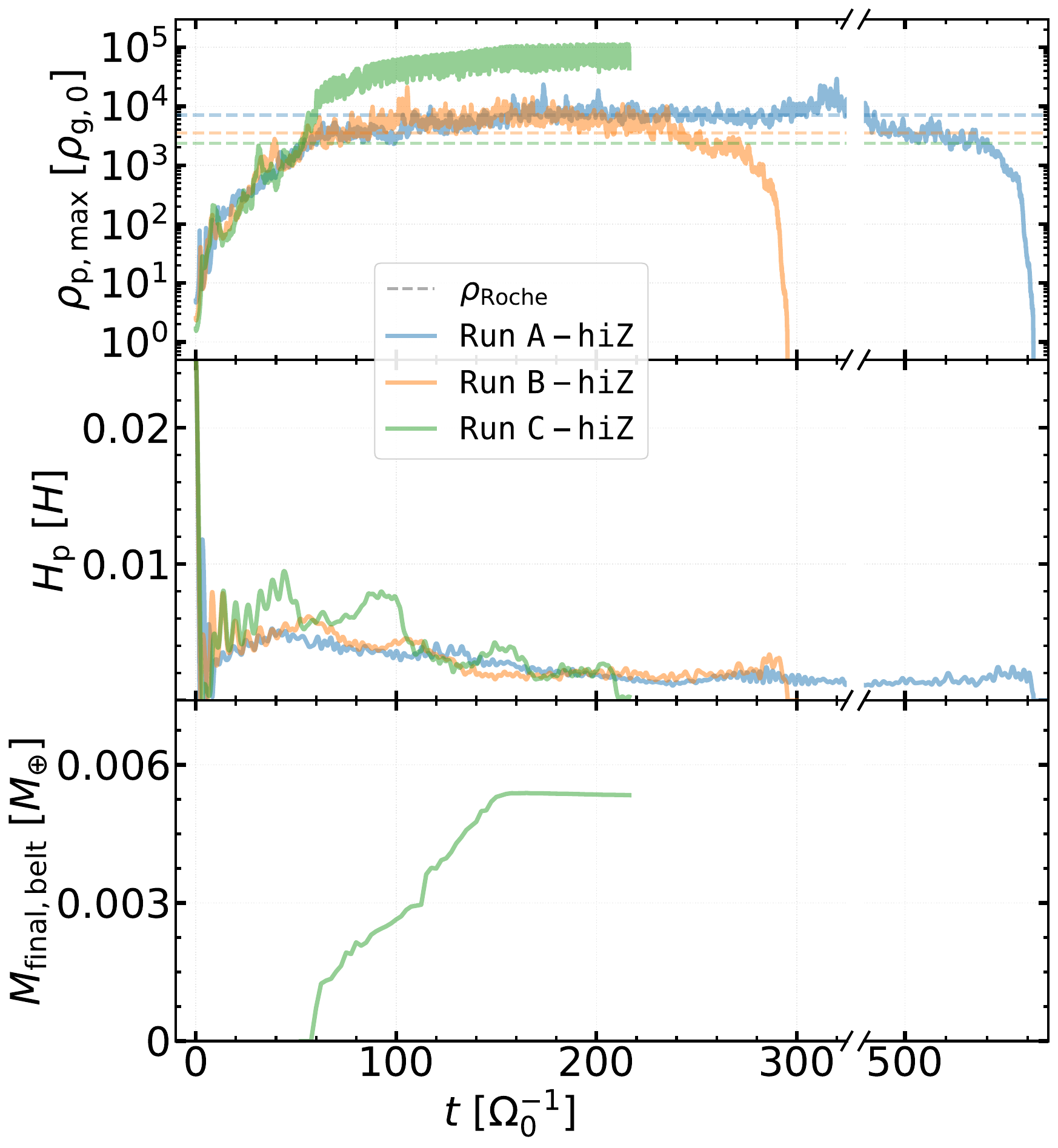}
  \caption{Same as Fig.~\ref{fig:par_stats} but for the higher metallicity runs with $Z > 0.01$ (height-integrated). See Table \ref{tab:paras} for parameters.
  \label{fig:par_stats_hiZ}}
\end{figure}

\begin{figure*}
  \centering
  \includegraphics[width=0.925\linewidth]{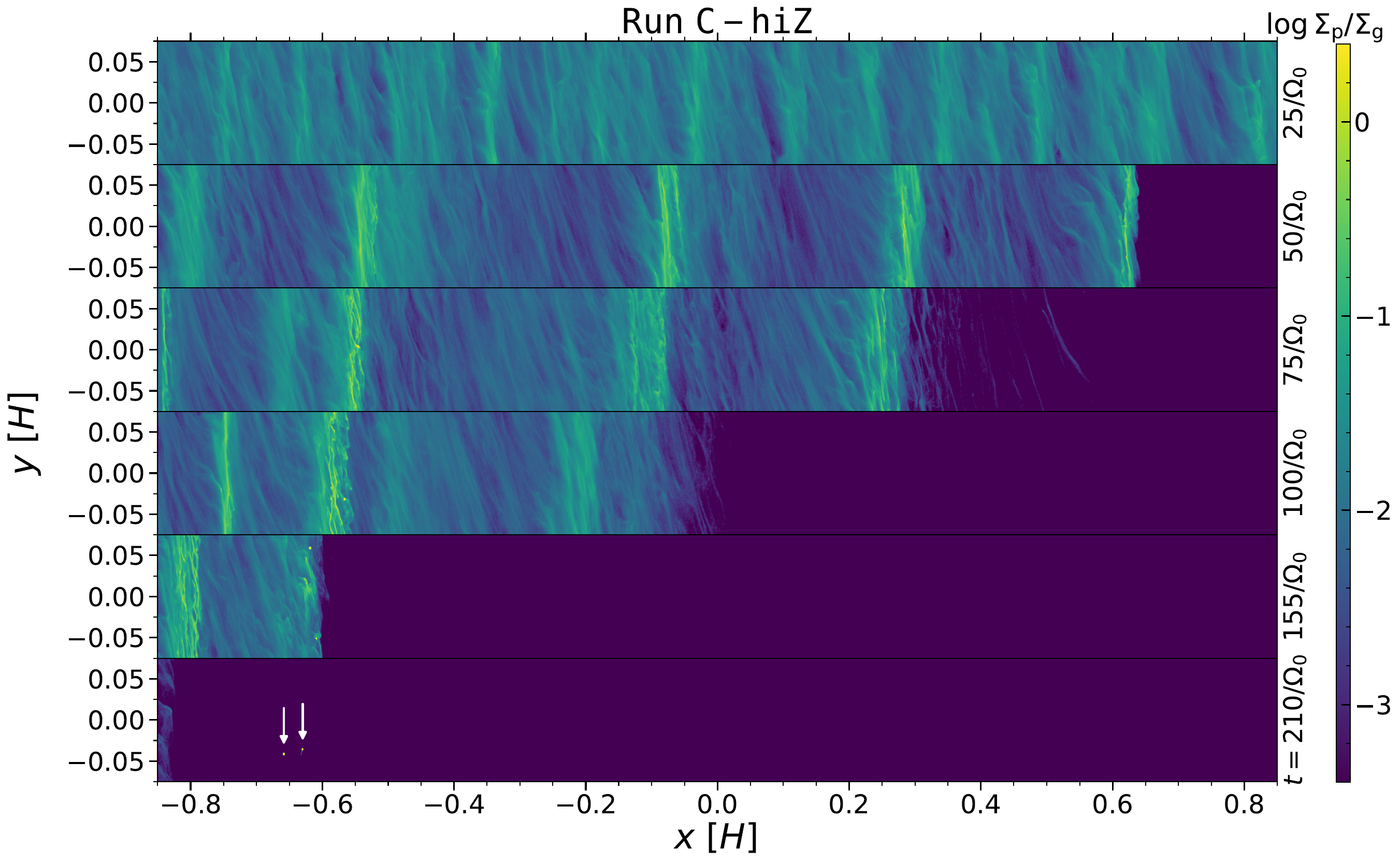}
  \caption{Same as Fig.~\ref{fig:new_snapshots1} but showing snapshot sequences from \texttt{Run C-hiZ}, which yields two surviving clumps at $x(H_{\rm g}) \approx -0.6$ to $-0.7$.  Simulation videos are available at \href{https://www.rixinli.com/cckb}{rixinli.com/CCKB}.
  \label{fig:new_snapshots3}}
\end{figure*}

\section{Results for Smooth Disk (No Bump) Simulations}
\label{sec:results}

We report on the outcomes of particle drift and clumping in smooth disk simulations as set up in Section \ref{sec:method}.  Section \ref{subsec:pf} measures the numbers and masses of planetesimals formed.  Section \ref{subsec:binarity} examines their binarity and obliquity. \rla{Section \ref{subsec:zoom_sim} expands upon all of these results with specially tailored simulations at higher spatial resolution.}

\subsection{Planetesimal Formation via Particle Drift in a Smooth Disk \rla{(Standard Resolution)}}
\label{subsec:pf}

Fig.~\ref{fig:par_stats} tracks how maximum particle densities $\rho_{\rm p,max}$ within the simulation box evolve with time for \rla{our standard resolution} \texttt{Runs A--D} (increasing $\mathcal{F}_{\rm gas}$ at fixed low $Z = 0.01$; see Table \ref{tab:paras}).  Also plotted are particle scale heights $H_{\rm p}$  evaluated as the standard deviation of vertical particle positions, and the mass of the belt in planetesimals, as extrapolated from persistent, self-bound clumps identified by \texttt{PLAN}.  Figs.~\ref{fig:new_snapshots1} and \ref{fig:new_snapshots2} provide some snapshots. 

At fixed $Z$, increasing $\mathcal{F}_{\rm gas}$ increases $M_{\rm init,belt}$, thereby lowering $Q$ and $\rho_{\rm R}/\rho_{\rm g,0}$; the density threshold for planetesimal formation is progressively eased from \texttt{Runs A} through \texttt{D}.  We find for \texttt{Runs B} to \texttt{D} that the maximum particle density $\rho_{\rm p,max}$ exceeds the Roche density $\rho_{\rm R}$ by increasing margins (Fig.~\ref{fig:par_stats}). Although \texttt{Runs A} and \texttt{B} do not feature gravitationally bound clumps, the streaming instability (SI) appears to be operating, as the maximum dust-to-gas density ratio reaches $\rho_{\rm p,max}/\rho_{\rm g,0} \sim 10^3$ and  persistent dust filaments (which do not clump further) are evident.  The \texttt{Run A} filament particle density falls short of $\rho_{\rm R}$ by a factor of several.

\texttt{Run C} is initialized with a total mass $M_{\rm init,belt} = 0.116 M_\oplus$.  Although $\rho_{\rm p,max}$ exceeds $\rho_{\rm R}$ in this run, it does so only intermittently by factors of a few, and the clumps formed are loosely bound, transient, and ultimately disrupted by tidal shear.  We are likely numerically under-resolving these clumps spatially and therefore under-estimating their self-gravity; it is conceivable that at least some of our clumps would remain bound at a higher grid resolution, \rla{an expectation we will confirm in Section \ref{subsec:zoom_sim}}.  Counting the unbound clumps as bound yields \rla{a higher estimate} for the total planetesimal belt mass of $M_{\rm final,belt} \simeq 1.32 \times 10^{-3} M_\oplus$ (listed in Table \ref{tab:paras} in parentheses), at the lower end of the range of mass estimates \rla{for the primordial CCKB (recall $M_{\rm CCKB} = 0.003 M_\oplus \pm 0.5$ dex as estimated in Section \ref{sec:intro}).}

\texttt{Run D} is initialized with $M_{\rm init,belt}=0.193 M_\oplus$. One bound clump forms and persists to the end of the simulation; in Fig.~\ref{fig:new_snapshots2}, we see that the clump has departed its natal dust filament. The clump has the mass equivalent of a binary having identical components of mean density $\rho_\bullet = 1$ g/cm$^{3}$ and radius $R_{\rm plan} \simeq 117$ km, comparable to the sizes of real-life CCKBOs (for justification of our interpretation of clumps as binaries, see \S\ref{subsec:binarity}). Extrapolated to the full $2\pi$ azimuthal extent of the CCKB, the estimated total mass in planetesimals is $M_{\rm final,belt} \simeq 1.43\times 10^{-3} M_\oplus$. A \rla{higher value of} $M_{\rm final,belt} \simeq 5.28 \times 10^{-3} M_\oplus$ is estimated by adding to the bound clump mass all the masses of the transient clumps that form and eventually disperse (but which persist in a higher resolution simulation; \rla{see Section \ref{subsec:zoom_sim}}). These simulated belt masses fall squarely in the range of possible \rla{primordial} CCKB masses. The least massive transient clump has as much mass as a binary composed of planetesimals of radius $R_{\rm plan} \sim 95$ km. 

The three \texttt{hiZ} runs listed in Table \ref{tab:paras} explore the role of disk metallicity $Z$. \texttt{Run A-hiZ} has identical input parameters to \texttt{Run A} except that $Z$ and by extension $M_{\rm init,belt}$ are increased by factors of 5. Despite the increase in solid mass in \texttt{Run A-hiZ}, no bound clumps form, transient or otherwise. The maximum particle density $\rho_{\rm p,max}$ exceeds the Roche density $\rho_{\rm R}$, but only by factors of $\sim$2--3, apparently not enough for gravitational collapse \citep{Gerbig2023}. For comparison, in \texttt{Run D} which yielded robust planetesimal formation, $\rho_{\rm p,max} > \rho_{\rm R}$ by a factor of $\sim$10. There is so little disk gas in \texttt{Runs A} and \texttt{A-hiZ} ($\mathcal{F}_{\rm gas} = 1\%$) that the Roche density is $\sim$7000$\times$ greater than the midplane gas density --- this is a high bar for concentration by the streaming instability to overcome (cf.~\citealt{LY21}). We can consider increasing $Z$ still further to force gravitational collapse, but such a scenario would only increase the initial belt mass $M_{\rm init,belt}$ beyond that of \texttt{Run D}, which we consider our ``most successful'' run insofar as it reproduces properties of the present-day CCKB with a minimum $M_{\rm init,belt}$. Moreover, we suspect that a metallicity $Z$ approaching 1 would potentially transform an order-unity fraction of the particles into planetesimals, which given the other parameters of \texttt{Run A-hiZ} would yield a final belt mass $M_{\rm final,belt}$ grossly exceeding that of the present-day CCKB. See Section \ref{sec:pure_GI} for further exploration of a disk dominated by solids.

\texttt{Runs B-hiZ} and \texttt{C-hiZ} are similarly analogous to their standard $Z$ counterparts.  Their initial belt masses $M_{\rm init,belt}$ are identical to that of the successful CCKB-forming \texttt{Run D}, but only \texttt{Run C-hiZ} succeeds in forming planetesimals (2 of them in the shearing box) that survive the duration of the simulation. Compared to \texttt{Run C} (which forms only transient clumps), \texttt{Run C-hiZ} increases $Z$ by a factor of just 5/3.

Taken together, our seven experiments indicate that if the CCKB formed from a ring of radius $r = 45$ au and a radial width of $5$ au (similar to the CCKB dimensions today), that ring would need to have an initial solid mass between that of \texttt{Run C} and \texttt{Run D}, i.e.~$M_{\rm init,belt} \simeq 0.15 M_\oplus$, roughly $70\times$ the mass in the CCKB today. \rla{This initial mass requirement might be lowered somewhat in higher-resolution simulations.}

\subsection{Binarity \rla{(Standard Resolution)}}
\label{subsec:binarity}

Here we examine, \rla{for our standard resolution runs,} the spin angular momenta of gravitationally collapsed clumps, both their scalar magnitudes to assess whether the clumps may fission into binaries (\S\ref{subsubsec:binarity1}) and their directions to compare with the inclinations of CCKB binaries (\S\ref{subsubsec:obliquities}). \rla{Results for higher resolution runs, which will show better agreement with observations, are discussed in Section \ref{subsec:zoom_sim}.}

\subsubsection{Spin Angular Momenta of Clumps}
\label{subsubsec:binarity1}

We measure the spin angular momentum $J$ of each clump measured relative to its barycenter. \texttt{PLAN} identifies 2 clumps in \texttt{Run C} (all transient), 7 clumps (6 transient, 1 persistent) in \texttt{Run D}, 6 clumps (all transient) in \texttt{Run B-HiZ}, and 7 clumps (5 transient, 2 persistent) in \texttt{Run C-HiZ}.

Figure~\ref{fig:RunCD_J_Ob} compares the measured $J$ values to their respective critical angular momenta 
\begin{equation}\label{eq:J_crit}
  \begin{aligned}
    J_{\rm crit} 
    &= 0.39 M_{\rm plan}^{5/3} G^{1/2} \left(\frac{3}{4\pi \rho_\bullet} \right)^{1/6}
  \end{aligned}
\end{equation}
above which a rotating, self-gravitating body of mass $M_{\rm p}$ and material density $\rho_\bullet$ cannot maintain a Jacobi ellipsoid shape \citep{Poincare1885}.  We find that $J/J_{\rm crit} > 1$ for every clump, implying that each cannot collapse into a single planetesimal, but would most likely fission into a binary \citep{Nesvorny2021}. Expressed in Hill units (top panel of Fig.~\ref{fig:RunCD_J_Ob}), the $J$'s are such that the binaries formed would extend to an order-unity fraction of their Hill spheres.

It is expected that $J/J_{\rm crit} > 1$ from the streaming instability. In the following order-of-magnitude argument, we drop all order-unity numerical factors. An SI-seeded clump would be expected to have an angular momentum 
\begin{equation}
  J \sim M_{\rm plan} \times \eta r \times \eta v_{\rm K} \,,
\end{equation}
where $\eta r$ and $\eta v_{\rm K} \sim \eta r \Omega_0$ are the characteristic length scale and velocity scale of the SI, and $M_{\rm plan}$ is the estimated clump mass based on the Roche density,\footnote{A more accurate estimate for $M_{\rm plan}$ would account for how clumps are created from filaments whose dimensions are only a fraction $f \sim 0.1$ of $\eta r$. Replacing $\eta$ with $f \eta$ throughout our order-of-magnitude derivation results in $f$ canceling out of our final answer for $J/J_{\rm crit}$.}
\begin{equation}
  M_{\rm plan} \sim \rho_{\rm R} \left(\eta r \right)^3 \,.
\end{equation}
Further approximating $J_{\rm crit} \sim M_{\rm plan}^{5/3} G^{1/2} \rho_\bullet^{-1/6}$, we have
\begin{equation}\label{eq:J_over_J_crit}
  \begin{aligned}
    \frac{J}{J_{\rm crit}} &\sim \left( \frac{\sqrt{G \rho_\bullet}}{\Omega_0} \right)^{1/3} \\
    &\sim 70 \left(\frac{\rho_\bullet}{1 \text{ g} \text{ cm}^{-3}} \right)^{1/6} \left(\frac{r}{45\text{ au}} \right)^{1/2} \,.
  \end{aligned}
\end{equation}
The empirical, precisely measured values of $J/J_{\rm crit}$ in Fig.~\ref{fig:RunCD_J_Ob} are of this order, though somewhat lower because not all of the particles that initially comprise a clump end up bound to the clump; some are tidally shorn off and carry away mass and spin angular momentum.  \citet{Nesvorny2021} measured that ``10\%–75\% of the original scaled angular momentum and 50\%–100\% of the original mass of a pebble clump is converted to the final binary system,'' \rla{ with roughly 10\% of clumps ``producing hierarchical systems with two or more large moons'' (see also \citealt{Robinson2020}).  We therefore expect our clumps to be mainly progenitors of binaries, with occasional higher-order systems.}

\begin{revfigure*}
  \centering
  \includegraphics[width=0.95\linewidth]{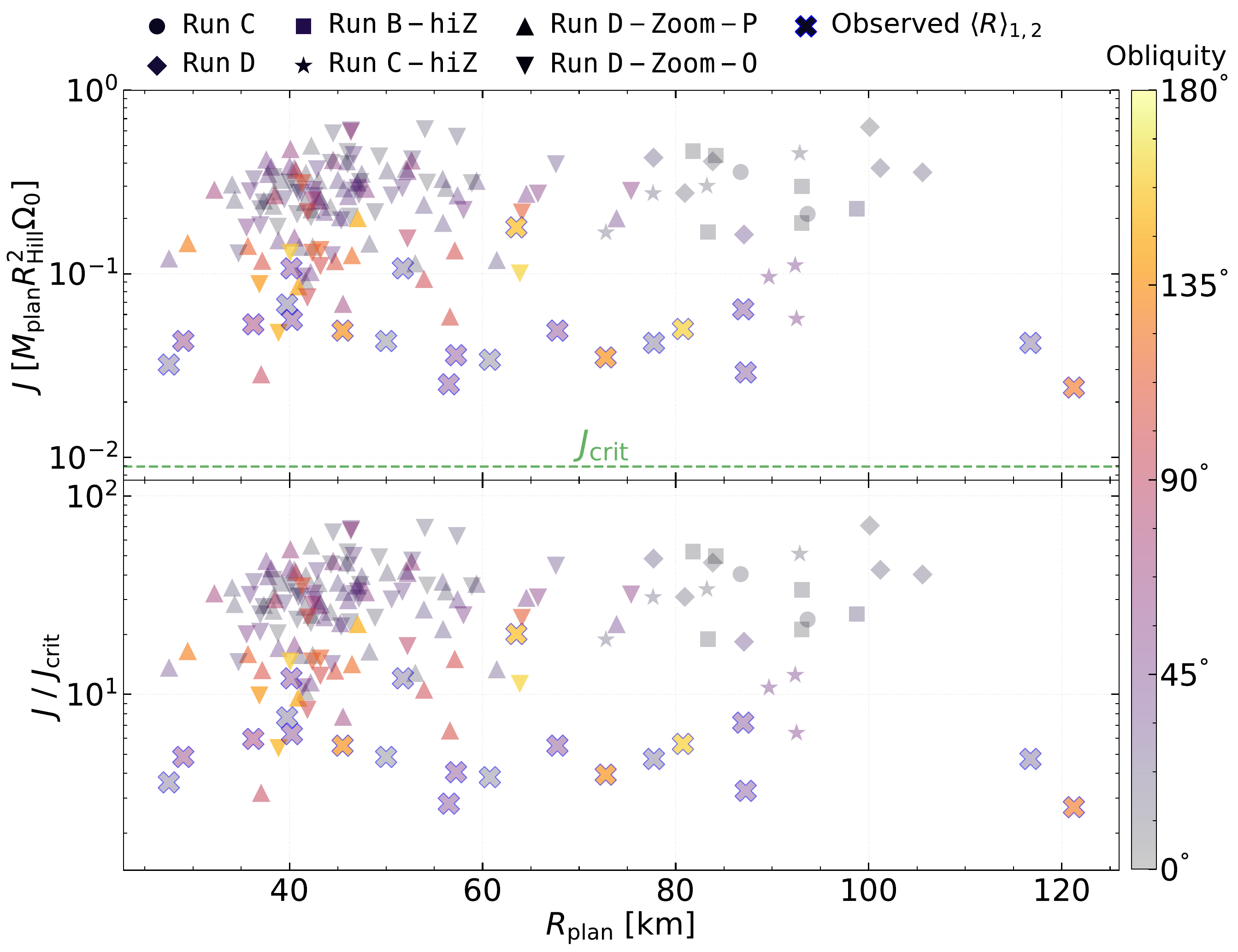}
  \caption{The spin angular momenta of clumps, both transient and persistent, identified in \texttt{Runs C}, \texttt{D}, \rla{\texttt{D-Zoom-P}, \texttt{D-Zoom-O}}, \texttt{B-hiZ}, and \texttt{C-hiZ}, in Hill units (top) and scaled by $J_{\rm crit}$ (bottom; eq. \ref{eq:J_crit}). \rla{Also shown are the orbital angular momenta of observed CCKB binaries with measured inclinations $i_{\rm bin}$ (defined as the angle between the binary orbit normal and the heliocentric orbit normal; data from \citealt{Nesvorny2021}). Data for clumps are color-coded by obliquities; $i_{\rm bin}$ for observed CCKB binaries uses the same color scale. For observed binaries, $R_{\rm plan}$ is the mean radius of the binary components (data from \citealt{JohnBinaryTable}). For simulated clumps}, $R_{\rm plan}$ is computed as the radius of a single planetesimal of material density $\rho_\bullet = 1$ g/cc, assuming a given clump forms two such planetesimals in an equal-mass binary. All simulation-based data in this figure are evaluated at the onset of a clump's formation, when clump properties are best compared to the predicted value (\ref{eq:J_over_J_crit}) from the SI (by contrast, Table \ref{tab:paras} evaluates $R_{\rm plan}$ at either the end of the simulation for a bound clump, or just before dispersal for a transient clump).  \rla{The simulated clumps show systematically higher $J$ than observed binaries; this discrepancy is expected to lessen as the clumps lose angular momentum during their collapse (see Section \ref{subsubsec:binarity1} and  \citealt{Nesvorny2021}).}
  \label{fig:RunCD_J_Ob}}
\end{revfigure*}

\begin{revfigure}
  \centering
  \includegraphics[width=\linewidth]{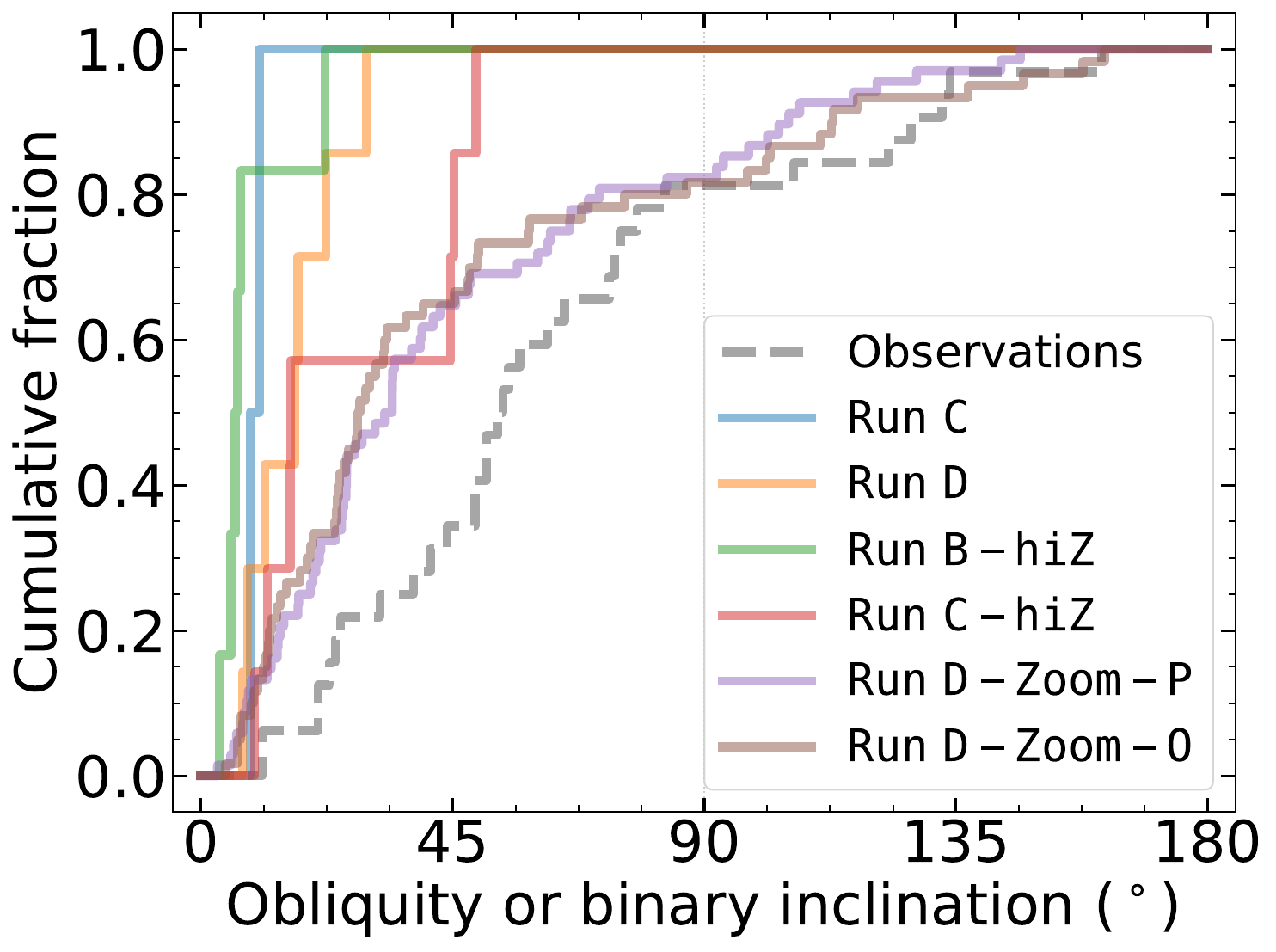}
  \caption{Comparison of the obliquity distributions of clumps (both transient and persistent) collected from \texttt{Runs C}, \texttt{D}, \texttt{B-hiZ}, \texttt{C-hiZ}, \rla{\texttt{D-Zoom-P}, and \texttt{D-Zoom-O}}, and the inclination distribution of observed CCKB binaries (grey dashed line; \citealt{Grundy2019}). \rla{The simulated obliquity distributions in the four standard-resolution runs are uncertain as they each rely on just a handful of clumps (2--7 depending on the run). The zoomed-in runs resolve 60--68 clumps having obliquities more closely resembling CCKB binary inclinations.}
  \label{fig:RunCD_Obliquity_CDF}}
\end{revfigure}

\begin{revfigure*}
  \centering
  \includegraphics[width=\linewidth]{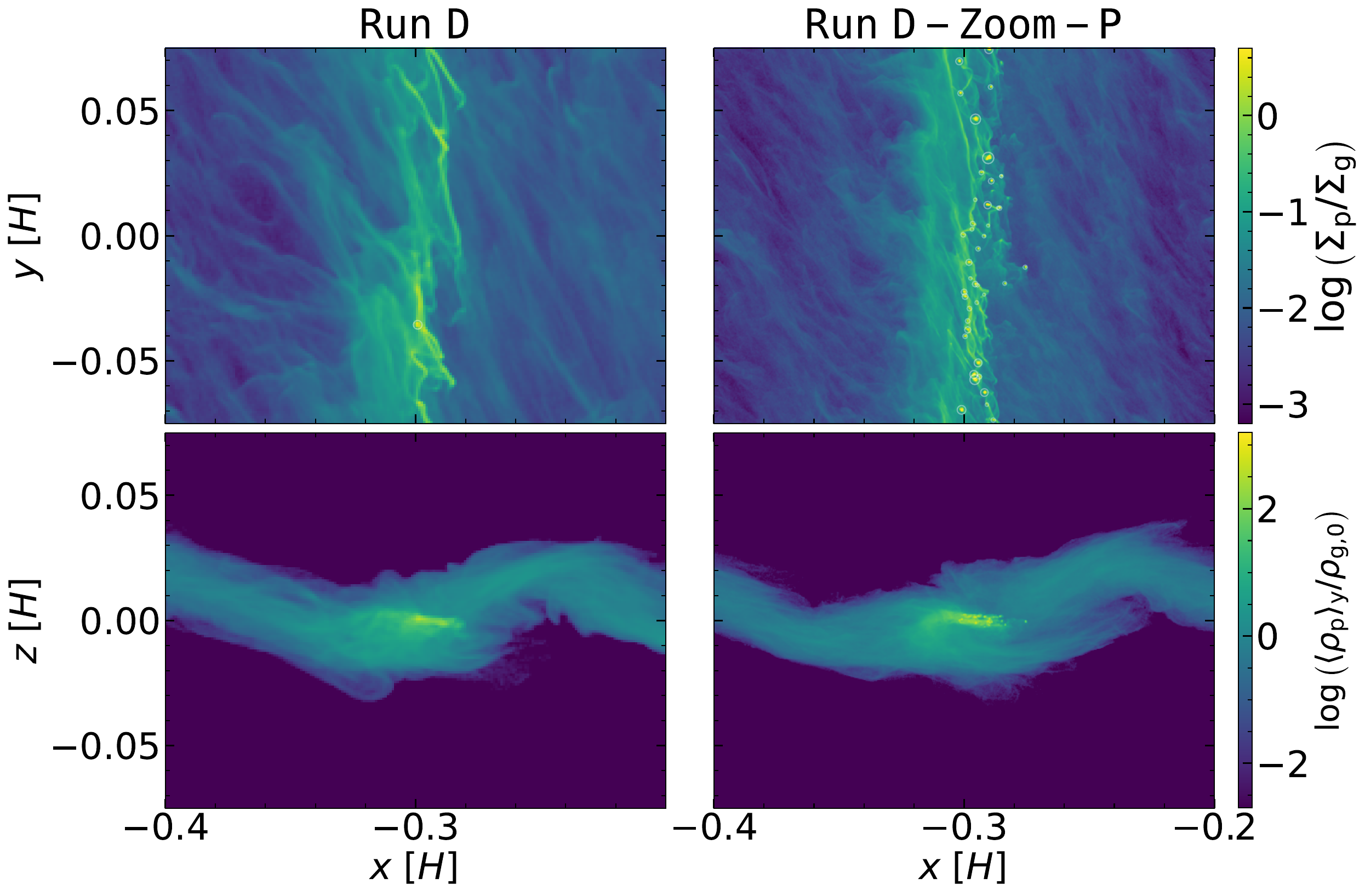}
  \caption{\rla{Comparison of snapshots of particle surface density (top row) and azimuthally averaged volume density (bottom row) between the original \texttt{Run D} ($\Delta x=H/1040$) and the zoomed-in simulation \texttt{Run D-Zoom-P} ($\Delta x=H/2560$) at $t=70 \Omega_0^{-1}$.  While the large-scale morphology of the dust filament and background particle layer are similar between the two runs, many more self-gravitating clumps are evident in the higher-resolution simulation. Clumps are marked in the top row by white circles indicating Hill spheres.}
  \label{fig:snapshot_RunD_vs_RunDzoomP}}
\end{revfigure*}

\subsubsection{Obliquities}
\label{subsubsec:obliquities}

Figure \ref{fig:RunCD_Obliquity_CDF} plots the obliquity distributions of clumps from \texttt{Runs C}, \texttt{D}, \texttt{B-hiZ}, and \texttt{C-hiZ} (see also Fig.~\ref{fig:RunCD_J_Ob} where obliquity is coded by color). Simulated clump spins --- from this small, resolution-limited sample \rla{(contrast with the higher-resolution results in Section \ref{subsec:zoom_sim})} --- are exclusively prograde. As can also be seen from Fig.~\ref{fig:RunCD_Obliquity_CDF}, observed CCKB binaries are 80\% prograde ($i_{\rm bin} < 90^\circ$), 20\% retrograde ($i_{\rm bin} > 90^\circ$; here $i_{\rm bin}$ is the angle between the binary orbit normal and the heliocentric orbit normal). Our \rla{standard-resolution} simulations do not recover a retrograde population, but suffer from small-number statistics. For example, in \texttt{Run C-hiZ} which exhibits 7 clumps, there is a $(0.8)^7 \simeq 21\%$ chance of drawing 7 prograde spins. \rla{See Section \ref{subsec:zoom_sim} for improved results at higher resolution.}

\citet{Nesvorny2019} achieved better agreement with the CCKB binary inclination distribution working in a different region of parameter space.  Their gas surface densities $\Sigma_{\rm g}$ were substantially larger --- $\sim$$8 \times$ greater than that of our most massive disk in \texttt{Run D} --- and their thresholds for gravitational instability were commensurately lower. Consequently, numerous clumps formed by gravitational collapse in their simulations; efficiencies of planetesimal formation ranged from 27--69\%, in contrast to our efficiencies of 0.7--2.8\%. Their order-unity efficiencies enabled the obliquity distribution to be well measured statistically, but are too large to be compatible with the CCKB --- their shearing-box outcomes, scaled the same way we scale ours, imply a final belt mass $M_{\rm final,belt}$ orders of magnitude above the present-day CCKB mass.

In general, the formation of retrograde clumps by gravitational collapse requires strong turbulence and vertical motions to tilt vorticity vectors (e.g.~\citealt{Jennings_Chiang2021}). \citet{Nesvorny2019} better resolved 3D motions and turbulent density enhancements as their grid cells were $\sim$$3\times$ smaller than ours. \rla{In the following Section \ref{subsec:zoom_sim}, we pursue higher-resolution simulations in an attempt to reproduce the observed retrograde population.}

\subsection{\rla{Zoomed-In Simulations (High Resolution)}}
\label{subsec:zoom_sim}

\rla{We conduct higher resolution, zoomed-in simulations of a single dust filament, basing them on a snapshot from \texttt{Run D} taken at $t = 60 \Omega_0^{-1}$ (before the maximum particle density has crossed the Roche density) of a filament located between $x=-0.4H$ and $x=-0.2H$ (see Fig. \ref{fig:new_snapshots2}).  This snapshot is used to construct \texttt{Runs D-Zoom-P} and \texttt{D-Zoom-O} as follows:}

\rla{
\begin{enumerate}
\item Gas quantities (density, momentum) are interpolated onto a new box of size $L_X'\times L_Y\times L_Z=(0.2\times 0.15\times 0.2)H^3$, with grid cells $\Delta x'= H/2560$ on a side (instead of our standard $H/1040$).  We apply shearing-periodic conditions for gas at the new radial boundaries; to prevent discontinuities, we smooth initial gas quantities, after interpolation, within cells located within $8\Delta x'$ of either radial boundary.
\item Each of the original \texttt{Run D} dust particles spawns $8$ new particles, each with $1/8$ the original particle mass but having identical velocities.  The new particles are initialized with positions shifted relative to the parent particle by $\Delta x'/2$ in random directions, ensuring that the overall dust density distribution remains unchanged on the new grid.  The number of dust super-particles is $N_{\rm p}=1.18\times 10^{8}$.
\item In \texttt{Run D-Zoom-P}, we apply shearing-periodic boundary conditions for particles in the radial direction, and run the simulation for an additional $10 \Omega_0^{-1}$ (from $t = 60\Omega_0^{-1}$ to $70 \Omega_0^{-1}$), long enough for clumps to form and gravitationally collapse.  The shearing-periodic BCs mimic the continuous background drift of particles in the original \texttt{Run D}, which brings new particles into and out of the filament for as long as there are particles at larger radii. Over the same interval of time in the original \texttt{Run D}, the particle mass inside the sampled region increases by 22\%. We mocked up this same increase in the zoomed-in run by manually and incrementally ramping up the mass of each dust particle.
\item  \texttt{Run D-Zoom-O} is an alternative experiment where we employ outflow boundary conditions in the radial direction for particles and run for an additional $20 \Omega_0^{-1}$ (from $t = 60\Omega_0^{-1}$ to $80 \Omega_0^{-1}$), long enough for the filament to encounter most of the dust particles in the new box that drift past, and for nascent clumps to depart the filament. 
\end{enumerate}
}

\rla{Figure \ref{fig:snapshot_RunD_vs_RunDzoomP} compares the simulated region of \texttt{Run D-Zoom-P} with the corresponding region in \texttt{Run D} at $t=70 \Omega_0^{-1}$.  The large-scale features match well. On smaller scales, the higher resolution, zoomed-in run yields substantially more clumps. Higher spatial resolution enables particle concentration and self-gravity to be modeled more accurately, so that gravitational collapse occurs at earlier times and on smaller scales.   The estimated total CCKB mass (extrapolated to the full $2\pi$ circumference of the belt) is $M_{\rm final,belt} \simeq 11.8\times10^{-3}M_\oplus$ in \texttt{Run D-Zoom-P}, and $8.9\times10^{-3}M_\oplus$ in \texttt{Run D-Zoom-O} (see Table \ref{tab:paras}).  Both values remain within the realm of possibility for the actual CCKB mass observed today.  The maximum, binary-component/mass-equivalent radii in the zoomed-in runs are 132 km and 118 km, comparable to that in \texttt{Run D}.  Most of the 68 clumps identified by \texttt{PLAN} in \texttt{Run D-Zoom-P}, and the 60 clumps in \texttt{Run D-Zoom-O}, have mass-equivalent radii of $\sim$50 km.}

\rla{To Figures \ref{fig:RunCD_J_Ob} and \ref{fig:RunCD_Obliquity_CDF} we have added  the obliquity distributions of the zoomed-in runs. Agreement with the observed inclination distribution of CCKB binaries is much improved over the standard resolution runs; retrograde spins are evident in the zoomed-in simulations, made possible by improved statistics and better resolution of 3D turbulent motions.}

\section{Ruling Out Other Formation Scenarios}
\label{sec:pure_GI}

We have shown in preceding sections that a belt mass in particles of $M_{\rm init,belt} \sim 10^{-1} M_\oplus$ can result in a planetesimal belt mass of $M_{\rm final,belt} \sim 10^{-3} M_\oplus$, of order the estimated mass in planetesimals in the Cold Classical Kuiper belt. The efficiency of this planetesimal formation process is low, $M_{\rm final,belt}/M_{\rm init,belt} \sim 1\%$, and motivates finding alternative formation scenarios with higher efficiencies. Here we consider whether the CCKB can coagulate with of order its current mass (an order-unity planetesimal formation efficiency) via gravitational collapse of particles, either in a gas-free setting (\S\ref{subsec:planring}) or in gas pressure bump  (\S\ref{subsec:bump_sims}). We will argue that neither scenario can reproduce the observed present-day properties of the CCKB.

\subsection{A Gas-Free, Self-Gravitating Particle Disk}
\label{subsec:planring}

We consider a disk of particles having a surface density equal to that of the CCKB at the time of its formation (see section \ref{sec:intro}):
\begin{align} \nonumber
  \Sigma_{\rm CCKB} &= \frac{M_{\rm CCKB}}{2\pi r \Delta r} \\&\simeq \frac{0.003 M_\oplus}{2\pi \times 45 \, {\rm au} \times 5 \, {\rm au}} \simeq 6 \times 10^{-5} \, {\rm g}/{\rm cm}^2 \,.
\end{align}
For such a disk to be susceptible to gravitational fragmentation, its Toomre $Q$ must be near unity:
\begin{equation}
  Q = \frac{c_{\rm p} \Omega_0}{\pi G \Sigma_{\rm CCKB}} \sim 1
\end{equation}
for particle velocity dispersion $c_{\rm p}$, orbital angular frequency $\Omega_0 \simeq 7 \times 10^{-10}$ rad/s, and gravitational constant $G$. The requisite velocity dispersion
\begin{equation}
  c_{\rm p} \sim 0.2 \, {\rm mm/s}
\end{equation}
appears too low to be realistically possible. For comparison, relative velocities in planetary rings (e.g.~around Saturn, Uranus) can approach such small values, as they are strongly damped by inelastic particle collisions that occur roughly once per orbit \citep{GT1978}. For the proto-CCKB's interparticle collision rate to approach the orbital frequency would require the belt's mass to be in micron-sized particles, small enough to be strongly perturbed by solar wind and radiation (not to mention the gravity of the planets). Moreover, relative velocities of order mm/s are too small compared to CCKBO Hill velocities of $\sim$1 m/s to produce any retrograde binary on Hill sphere scales \citep{Schlichting2008,Schlichting2008a}.

\begingroup 
\setlength{\medmuskip}{0mu} 
\begin{deluxetable*}{cccccccccc}[ht]
  \tablecaption{Turbulent Bump Simulation Parameters}\label{tab:paras2}
  \tablecolumns{10}
  \tablehead{
    \colhead{(1)} &
    \colhead{(2)} &
    \colhead{(3)} &
    \colhead{(4)} &
    \colhead{(5)} &
    \colhead{(6)} &
    \colhead{(7)} &
    \colhead{(8)} &
    \colhead{(9)} &
    \colhead{(10)} \\
    \colhead{\texttt{Run}} &
    \colhead{$\mathcal{F}_{\rm gas}$} &
    \colhead{$Z$} &
    \colhead{$Q$} &
    \colhead{$\uptau_{\rm s,1mm}$} &
    \colhead{$\langle \delta u_x^2 \rangle$} &
    \colhead{$\langle \delta u_z^2 \rangle$} &
    \colhead{$M_{\rm init,belt}$} &
    \colhead{$M_{\rm final,belt}$} &
    \colhead{$R_{\rm plan}$} \\
    \colhead{} &
    \colhead{} &
    \colhead{} &
    \colhead{} &
    \colhead{} &
    \colhead{[$10^{-4} c_{\rm s}^2$]} &
    \colhead{[$10^{-4} c_{\rm s}^2$]} &
    \colhead{[$M_{\oplus}$]} &
    \colhead{[$10^{-3}M_{\oplus}$]} &
    \colhead{[km]}
  }
  \startdata
  \hline
  \texttt{Bu-1-nSG} &$2\%$ &\rlr{$0.0017$} &$\infty$ &$1.08$ &\rlr{$0.83$} &\rlr{$3.65$} &\rlr{$0.003$} &$0$    &…  \\
  \texttt{Bu-1}     &$2\%$ &\rlr{$0.0017$} &$626.4$  &$1.08$ &\rlr{$0.82$} &\rlr{$3.64$} &\rlr{$0.003$} &$0$    &…  \\
  \texttt{Bu-10}    &$2\%$ &\rlr{$0.0165$} &$626.4$  &$1.08$ &\rlr{$0.83$} &\rlr{$3.65$} &\rlr{$0.03$}  &$0$    &…  \\
  \texttt{Bu-100}   &$2\%$ &\rlr{$0.165$}  &$626.4$  &$1.08$ &\rlr{$0.88$} &\rlr{$3.40$} &\rlr{$0.3$}   &\rlr{$92.0$} &\rlr{$2504$} \\
  \enddata
  \tablecomments{Columns: 
  (1) run name;
  (2) ratio of gas surface density $\Sigma_{\rm g}$ relative to MMSN at $r = 45$ au ($\Sigma_{\rm g,MMSN} = 7.3$ g/cm$^2$); 
  (3) height-integrated metallicity $\Sigma_{\rm p}/\Sigma_{\rm g}$; 
  (4) Toomre $Q$ for the gas disk; 
  (5) dimensionless stopping time for dust particles of size $a = 1$ mm;
  (6) radial velocity dispersion ($\delta u_x = u_x$ assuming the background radial velocity is zero), spatially averaged over dust-free regions and time-averaged over the last $100/\Omega_0$ ($50/\Omega_0$ for \texttt{Run B-100} because of that run's shorter duration);
  (7) vertical velocity dispersion ($\delta u_z = u_z$ assuming the background vertical velocity is zero), space and time-averaged as in (6);
  (8) initial ``belt''  mass in particles, computed by scaling our shearing box to an annulus of radial width $5$ au centered at $r = 45$ au;
  (9) final ``belt'' mass (scaled as in (8)) in persistent clumps formed from the gravitational collapse of particles; 
  (10) planetesimal radius computed from the most massive persistent clump, assuming the clump forms an equal-mass binary. The listed radius is of one binary component, of mean density $\rho_\bullet = 1$ g/cc.  Simulation videos are available at \href{https://www.rixinli.com/cckb}{rixinli.com/CCKB}.
  }
\end{deluxetable*}
\endgroup

\begin{newfigure}
  \centering
  \includegraphics[width=\linewidth]{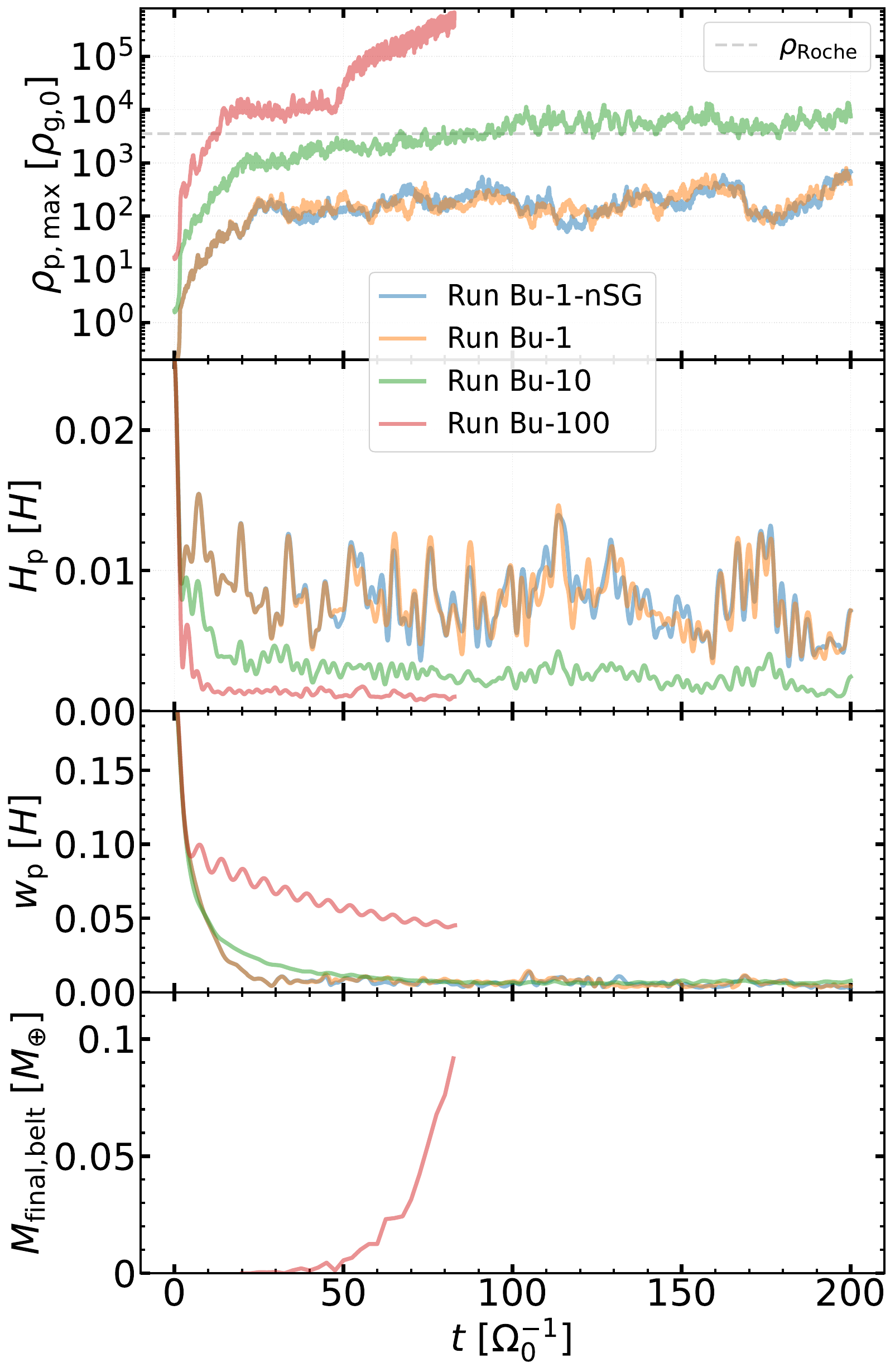}
  \caption{Similar to Fig. \ref{fig:par_stats} but for turbulent gas bump simulations (Table \ref{tab:paras2}).  The third panel from the top shows the Gaussian-fitted radial width of the dust ring. Only \texttt{Run Bu-100} forms planetesimals, but the formation efficiency (from wholesale gravitational collapse) is too large, resulting in a planetesimal belt nearly two orders of magnitude more massive than the CCKB today (bottom panel). 
  \label{fig:par_stats_bump}}
\end{newfigure}

\begin{newfigure*}
  \centering
  \includegraphics[width=\linewidth]{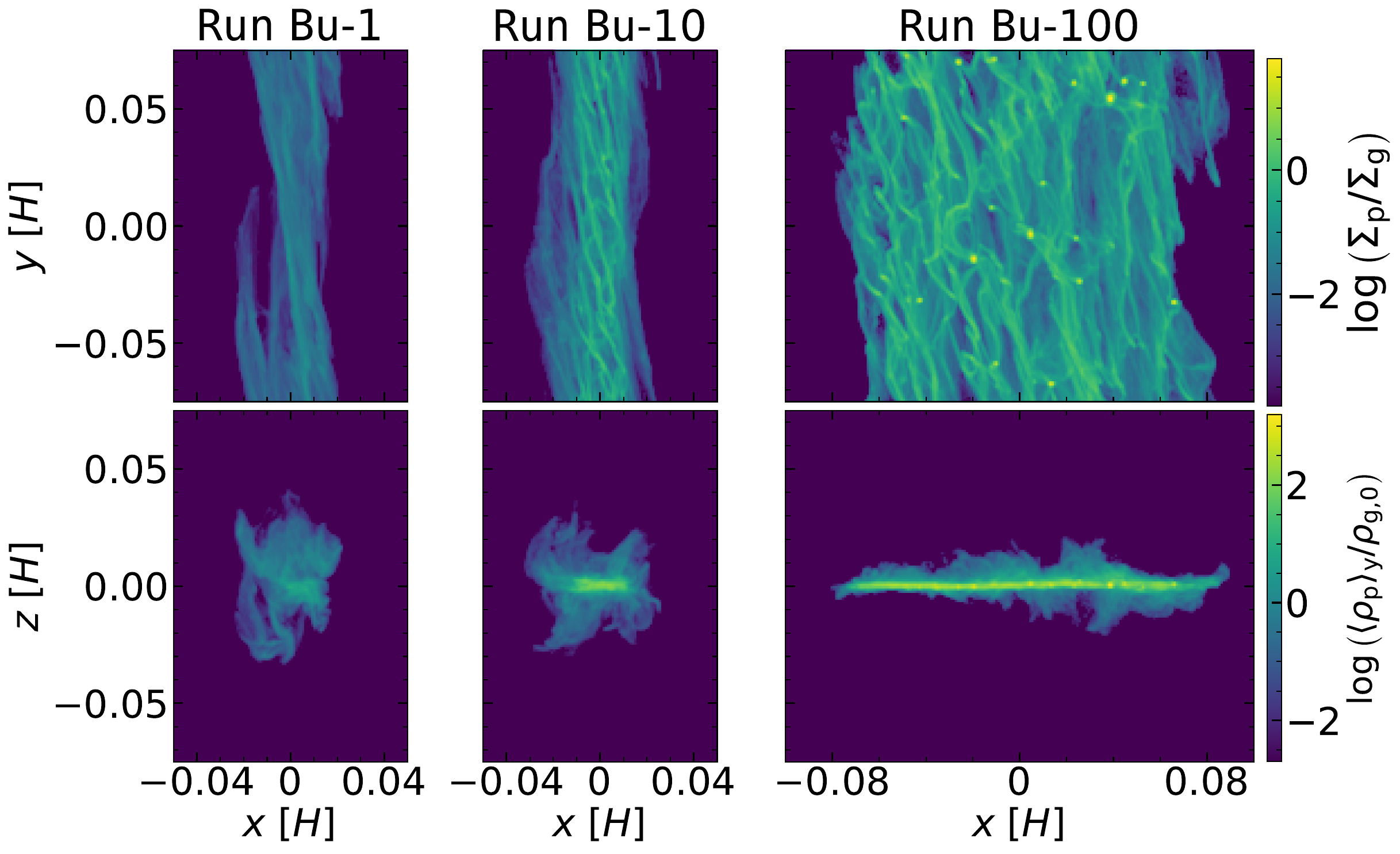}
  \caption{Snapshots of particle surface density (top row) and azimuthally averaged volume density (bottom row) for  turbulent gas bump simulations (\texttt{Runs Bu}) at $t=100\Omega_0^{-1}$ \rlr{($80\Omega_0^{-1}$ for \texttt{Run Bu-100})}. Either no planetesimals form (left and middle columns) or too many form (right), with no middle ground to match the present-day properties of the CCKB. For quantitative details, see Fig.~\ref{fig:par_stats_bump} and Table \ref{tab:paras2}.  Simulation videos are available at \href{https://www.rixinli.com/cckb}{rixinli.com/CCKB}.
  \label{fig:snapshots_bump}}
\end{newfigure*}

\subsection{Planetesimal Formation in a Gas Pressure Bump}
\label{subsec:bump_sims}

The efficiency of planetesimal formation can conceivably be increased by collecting dust particles in a gas pressure bump (see section \ref{sec:intro}). Following, e.g., \citet{Dullemond2018}, we envision a gas pressure bump centered at $r$ and of vertical scale height $H$ and radial width $w \geqslant H$, inside of which is a dust particle ring having a characteristic vertical thickness $H_{\rm p} \leqslant H$ and radial width $w_{\rm p} \leqslant w$. Assuming that the gas is stirred by turbulence (of unspecified origin) having a dimensionless diffusivity $\alpha \ll \uptau_{\rm s}$ (for dimensionless particle stopping time $\uptau_{\rm s}$), and ignoring the feedback of dust on gas, we have
\begin{align}
  H_{\rm p} \sim H \sqrt{ \frac{\alpha}{\uptau_{\rm s}}} \label{eq:hp_alpha}\\
  w_{\rm p} \sim w \sqrt{ \frac{\alpha}{\uptau_{\rm s}}} \label{eq:wp_alpha}
\end{align}
(see also \citealt{Youdin2007}). For small enough $\alpha/\uptau_{\rm s}$, and by extension small enough $H_{\rm p}$ and $w_{\rm p}$, one can imagine dust concentrates enough to trigger the streaming and/or gravitational instability. A potential obstacle to the SI is that it needs a radial pressure gradient which pressure bumps lack at their centers, by construction (e.g.~\citealt{Xu_Bai_2022}).

To test these ideas, we conduct a series of pressure bump simulations differing from our smooth disk simulations as follows:
\begin{enumerate}
\item Gas azimuthal velocities are forced to conform to those from a gas pressure bump having a Gaussian surface density in radius $x$:
\begin{equation}\label{eq:G_bump}
  \Sigma_{\rm g}(x) = \Sigma_{\rm g0} \left[ 1 + \mathcal{A} \exp\left( -\frac{x^2}{2 w^2} \right) \right], 
\end{equation}
for constant $\Sigma_{\rm g0} = \mathcal{F}_{\rm gas} \Sigma_{\rm g,MMSN}$, bump amplitude $\mathcal{A} = 0.25$, and bump radial width $w = H$. The corresponding background gas velocities are super-Keplerian at $x < 0$ and sub-Keplerian at $x > 0$, causing particles to drift toward $x = 0$. Manually enforcing the velocity profile at every timestep (via Newtonian relaxation with a timescale of $0.1\Omega_0^{-1}$) accounts for the bump's radial pressure gradient (we set the global pressure gradient parameter $\Pi = 0$), and maintains the bump against momentum feedback from dust.
\item To reconcile the velocity difference (super-Keplerian vs.~sub-Keplerian) at the shearing-periodic radial boundaries, we smooth the gas density so that its radial derivative is zero at either radial boundary, with corresponding adjustments to the azimuthal gas velocity.  See Appendix \ref{app:smooth} for details. 
\item Following \citet[][their section 2.3 and appendix]{Lim2024}, we force the gas to be turbulent by introducing an extra force density to the right-hand side of the gas momentum equation (\ref{eq:gasmom}):
\begin{equation}
    \rho_{\rm g} \bm{f}_{\rm turb} = \rho_{\rm g} \Lambda \left(\bm{\nabla} \times \bm{A} \right),
\end{equation}
where $\bm{A}$ is a vector potential with randomized phases 
(defined by equation A1 of \citealt{Lim2024}) and $\Lambda$ is a forcing amplitude. 
\footnote{\rla{\citet{Lim2024} varied the forcing phase at intervals of $10^{-3} \Omega_0^{-1}$ and found no significant difference for a slower interval of $10^{-2} \Omega_0^{-1}$ (Lim 2025, personal communication).  We experimented with still longer intervals of $10^{-1} \Omega_0^{-1}$ and $\Omega_0^{-1}$, and found that the forcing became so coherent that dust particles were expelled from the simulation domain.}}
We set $\Lambda = 10^{-4} c_{\rm s}^2$ and measure gas velocity dispersions in the radial and vertical directions of $\langle \delta u_x^2\rangle \simeq 0.8 \times 10^{-4} c_{\rm s}^2$ and $\langle \delta u_z^2\rangle \simeq 3.6 \times 10^{-4} c_{\rm s}^2$, where $\delta u$ is the random gas velocity and $\langle … \rangle$ is a spatial and time average over a volume of dust-free gas (for details see Table \ref{tab:paras2}).  Modulo the mild anisotropy (which arises because $\bm{A}$ depends on our unequal shearing box lengths), these velocity dispersions translate approximately to a  Shakura-Sunyaev dimensionless diffusivity of $\alpha \sim \langle \delta u^2 \rangle / c_{\rm s}^2 \sim 10^{-4}$, similar to the diffusivities inferred from DSHARP and exoALMA observations of dust rings within gas pressure bumps \citep{Dullemond2018,Stadler2025}.
\item The box size is $L_X \times L_Y \times L_Z =  (0.4 \times 0.15 \times 0.25) H^3$. We can afford a smaller radial length $L_X$ compared to the smooth disk simulations because dust concentrates strongly radially.  The vertical extent $L_Z$ is slightly larger to accommodate the thicker particle layer from  externally imposed turbulence.  Grid cells remain at our standard value of $\Delta x = H/1040$ on a side.
\item The number of Lagrangian dust super-particles is $N_{\rm p}\approx 1.35\times 10^{7}$ (from equation \ref{eq:N_p} with $n_{\rm p}=4$).  As all particles drift toward the gas bump center, the effective particle resolution increases. Dust particles are initialized the same way as in our smooth disk simulations.
\item We run for $200\Omega_0^{-1}$, more than long enough for dust to relax into a quasi-steady equilibrium balancing drift toward to the bump center and turbulent diffusion. We end \texttt{Run Bu-100} at \rlr{$82.8\,\Omega_0^{-1}$} because the strong dust overdensities in that simulation render the timesteps prohibitively short (see section 4.1.1 in \citealt{LY21}).
\end{enumerate}

Parameters and outcomes for the bump simulations (\texttt{Runs Bu}) are summarized in Table \ref{tab:paras2}. All four runs have the same gas bump properties ($\mathcal{F}_{\rm gas} = 2\%$, $w = H$, $\mathcal{A} = 0.25$, $\Lambda = 10^{-4}c_{\rm s}^2$) and particle stopping times ($\uptau_{\rm s,1mm} = 1.08$ for a fixed particle radius $a = 1$ mm and our assumed $\mathcal{F}_{\rm gas}$). \texttt{Runs Bu-1} through \texttt{Bu-100} differ only in their height-integrated metallicities $Z$, which survey initial belt masses $M_{\rm init,belt}$ ranging from $1\times$ to $100\times$ the assumed primordial CCKB mass, \rlr{$M_{\rm CCKB} = 0.003 M_\oplus$}.  \rla{Our $\alpha/\uptau_{\rm s}$ values (evaluated for millimeter-sized grains) are of order $10^{-4}$, three orders of magnitude smaller than those inferred for dust rings observed by ALMA \citep{Zagaria2023, Lee2024}. The latter estimates apply to gas-rich disks having $\mathcal{F}_{\rm gas} > 100\%$. For the gas-depleted conditions of interest here,  $\alpha/\uptau_{\rm s}$ is expected to be much smaller, given that $\uptau_{\rm s}\propto 1/\Sigma_{\rm g}$, and assuming $\alpha$ is independent of $\Sigma_{\rm g}$.}
 
\texttt{Run Bu-1} serves as a reference. As can be seen in Figure \ref{fig:par_stats_bump}, particle self-gravity is negligible for \texttt{Run Bu-1}, whose results appear statistically identical to those of \texttt{Run Bu-1-nSG} where self-gravity is turned off. No planetesimals form in \texttt{Run Bu-1}. Particle vertical scale heights $H_{\rm p}$ (evaluated as the standard deviation of vertical particle positions) and radial widths $w_{\rm p}$ (fitted to Gaussians) range from $\sim$0.005--0.01$H$, consistent to within a factor of 2 with the predictions of equations (\ref{eq:hp_alpha})--(\ref{eq:wp_alpha}) for $\alpha \sim 10^{-4}$ and $\uptau_{\rm s} = 1$. Interestingly, though the maximum particle density $\rho_{\rm p,max}$ falls well short of the Roche density, it is still $\sim$$100\times$ larger than the midplane gas density $\rho_{\rm g,0}$ (Fig.~\ref{fig:par_stats_bump} top panel). Particle densities vary strongly and stochastically from the imposed turbulent stirring. Even at the center of the gas bump to which all particles are attracted, particle overdensities can give way to voids, and back again, over dynamical times (see our simulation videos at \href{https://www.rixinli.com/cckb}{rixinli.com/CCKB}). Heavy particle mass loading, i.e.~strong feedback of particles on gas, weakens the ability of the latter to turbulently stir the former, thereby allowing particles to collect into especially small volumes, albeit transiently. \citet{Xu_Bai_2022} document similar effects in their (non-self-gravitating) simulations; see, e.g., the difference between their Z2 and  corresponding no-feedback Z0 curves in the top panel of their Fig.~2.

Snapshots of \texttt{Runs Bu-1} through \texttt{Bu-100} are given in Figure \ref{fig:snapshots_bump}. In all of the runs, solid particles collect into the centermost regions of the bump, occupying just a few percent of the total bump radial width where gas-particle relative streaming velocities are too low to power the streaming instability effectively. Only \texttt{Run Bu-100} forms planetesimals, but too many of them: an initial belt mass in particles of \rlr{$0.3 M_\oplus$} is converted with order-unity efficiency into planetesimals of total mass \rlr{$0.092 M_\oplus$}, $50\times$ too large compared to the present-day mass of the CCKB. The planetesimal mass is actually still growing at the time the simulation is stopped (Fig.~\ref{fig:par_stats_bump} bottom panel). The largest planetesimal formed is also too large compared to real-life CCKBOs, with a radius exceeding 2000 km (Table \ref{tab:paras2}). Planetesimal formation in \texttt{Run Bu-100} appears to proceed by wholesale gravitational collapse of particles within an especially thin, dense layer at the midplane (see the bottom right panel of Fig.~\ref{fig:snapshots_bump}, and the small $H_{\rm p}$ values in Fig.~\ref{fig:par_stats_bump}).

In sum, our experiments indicate that planetesimal formation in a turbulent gas pressure bump requires extreme particle mass loading to trigger gravitational instability. For a depleted late-stage background gas disk ($\mathcal{F}_{\rm gas} = 2\%$) containing mm-sized particles with order-unity Stokes numbers, and an ALMA-motivated turbulent diffusivity $\alpha \sim 10^{-4}$, either no planetesimals form for \rlr{$M_{\rm init,belt}\lesssim 0.1 M_\oplus$}, or too massive belts containing too large bodies form, with no middle ground to match the primordial CCKB mass of \rlr{$\sim$$0.003 M_\oplus$}. 

One can imagine adjusting $\alpha$ down to allow for gravitational collapse of smaller belt masses. This possibility strikes us as ad hoc and fine-tuned. When we simulate a strictly laminar gas pressure bump ($\Lambda = \alpha = 0$), we find planetesimal formation to be too effective --- all of the particles migrate unhindered to the bump center and agglomerate into a single gargantuan body (data not shown; see also \citealt{Lee2022}).

\section{Summary}
\label{sec:summary}

Cold Classical Kuiper belt objects (CCKBOs), solid bodies $\sim$100 km in diameter circling the Sun at 42--47 au, might be the Solar System's sole surviving first-generation planetesimals. Their dynamically cold orbits, fragile binarity, and distinctive surface colors suggest they have not been disturbed, collisionally or gravitationally, since birth.

We have investigated the formation circumstances of CCKBOs, using a 3D stratified shearing box to simulate how mm-sized particles embedded in circumstellar gas  clump and self-gravitate into larger bodies. Our main challenge was to understand how the CCKB's low mass at birth, $M_{\rm CCKB} = 0.003 M_\oplus \pm 0.5$ dex, may be reproduced by an in-situ, self-gravitational collapse scenario. As the corresponding solid disk surface density is three orders of magnitude lower than in typical, median-age protoplanetary disks, the CCKB likely formed at the tail end of the solar nebula's life, when primordial solids and gas had nearly but not completely dissipated.

A way for a small amount of solids to achieve the super-Roche densities required for self-gravitational collapse is to collect them by gas drag within a local pressure maximum. Bright dust rings are famously observed at sub-mm wavelengths in young gas disks and have been confirmed to be situated inside overdense gas rings, a.k.a.~gas pressure ``bumps'' which serve as dust traps (\citealt{Stadler2025}, and references therein). For gas bumps with multiple potential causes  (planets, magnetic instabilities, infall from the parent molecular cloud, ...), several groups have concluded that dust trapping can promote planetesimal formation (e.g.~\citealt{Stammler2019}; \citealt{Carrera2021}; \citealt{Carrera2022}; \citealt{Xu_Bai_2022}; \citealt{Zagaria2023}; \citealt{Zhao2025}). The most realistic of these studies, involving 3D numerical simulations of dust and gas, reveal that the streaming instability (SI; \citealt{Youdin_Goodman_2005}) does not necessarily play a strong role in dust clumping, as pressure bumps weaken the radial pressure gradient on which the SI relies. To our knowledge, all published gas bump scenarios assume dust masses orders of magnitude in excess of $M_{\rm CCKB}$, with typical dust ring masses $\gtrsim 1 M_\oplus$.

Could the CCKB have formed within a gas bump/dust trap, scaled to a low mass? As natural as this idea seems, we have been unable to make it work. We have numerically simulated dust in 3D inside manually enforced gas pressure bumps, accounting for self-gravity and varying degrees of turbulence. If the gas is not turbulent (apart from being stirred by interactions with dust), solid particles drift unimpeded into the gas bump's center and agglomerate into a single body --- not the hundreds of bodies present in the CCKB. If we introduce gas turbulence with a momentum diffusivity $\alpha \sim 10^{-4}$ similar to that inferred from protoplanetary disk observations \citep{Dullemond2018,Stadler2025}, then mm-sized particles are kept diffused within the bump, but fail to clump and form planetesimals when their collective mass $M_{\rm init,belt} \leq 10 M_{\rm CCKB}$. At a larger initial belt mass of $M_{\rm init,belt} = 100 M_{\rm CCKB}$, we have the opposite problem --- more than half of the particles gravitationally collapse into a too-massive belt filled with super-Plutos. It appears that planetesimal formation in gas pressure bumps, with and without turbulence, leaves no middle ground to conceive the low-mass CCKB (for simulation videos, see \href{https://www.rixinli.com/cckb}{rixinli.com/CCKB}).

Instead of a dust trap, our numerical experiments favor a drift-and-form scenario. In a smooth (no bump) disk, dust particles encounter the usual nebular headwind to migrate radially inward. For a wide range of particle Stokes numbers $\uptau_{\rm s} \lesssim 1$ (particle stopping times $\lesssim$ the orbital period/$2\pi$) and disk metallicities $Z$ (height-integrated dust-to-gas ratios) $\gtrsim 0.005$-$0.01$, particles may clump from the streaming instability (e.g. \citealt{LY21}; \citealt{Lim2025}). The strongest overdensities collapse further into self-gravitating planetesimals, which largely decouple from the background drift because they are more massive and affected less by the headwind.

In a drift-and-form scenario, CCKBO formation may be inefficient, as for every gram of planetesimal formed at a given orbital radius, there are many more grams of dust that drift past (a situation not unlike pebble accretion; e.g.~\citealt{Lin2018}). We found that in a late-stage depleted disk having $\mathcal{F}_{\rm gas} \sim 2$--5\% of the gas contained in the minimum-mass solar nebula, a collection of mm-sized particles of mass $M_{\rm init,belt} \simeq 0.1$--$0.2 M_\oplus \sim 50$--$100 M_{\rm CCKB}$ may drift out of the 42--47 au heliocentric annulus, leaving behind a total mass in planetesimals of $M_{\rm final,belt} \simeq (1$--$8) \times 10^{-3} M_\oplus$, \rla{in good agreement with the estimated birth mass of the CCKB.} The corresponding $Z \simeq 0.01$--0.03 and $\uptau_{\rm s} \simeq 0.4$--1. \rla{The required initial belt mass cited above may be overestimated as it derives from simulations whose spatial resolution could be improved. The $\sim$0.1--0.2 $M_\oplus$ of solids that drift out of the CCKB region would encounter the giant planets and either be accreted by them or gravitationally scattered into the dynamically hotter portions of the Kuiper belt and the Oort Cloud.}

This drift-and-form scenario reproduces not only the present-day mass of the CCKB, but also the typical sizes of CCKBOs and their binarity. In our simulations, each bound clump has enough mass and spin angular momentum to fission into an equal-mass binary composed of planetesimals of radius $R_{\rm plan} \simeq 40$--100 km. \rla{In our highest-resolution simulations, the obliquity distribution of the clumps resembled the inclination distribution of CCKB binaries \citep{Grundy2019}; in particular we recovered a population of retrograde clumps.}

\rla{Perhaps the most pressing problem posed by our proposed solution is articulating a  ``backstory'' for the background particle drift.} It takes only $\sim$10$^4$ yr for our hypothesized sheet of dust particles in leftover disk gas to traverse the CCKB region. This is a short interval of time compared to disk ages of $\sim$$10^6$ yr, and begs the question what else happened in this region, and in the exterior regions that fed it, while the disk was still present. Perhaps there were previous drift-and-form episodes that populated CCKB space, and these earlier generations of planetesimals were later purged during the chaos that created the dynamically hot components of the Kuiper belt.\footnote{\rla{See, e.g., \citet{Lau2025} for an attempt to form the Kuiper belt in a late-stage disk, but note that their 1D (radial) treatment of the solar nebula relies on prescriptions for planetesimal formation that do not capture the outcomes of our 3D simulations.}} If nebular gas decays linearly with time --- as would be the case for a constant mass-loss rate set by a photoevaporative wind from the gas disk surface --- our modeled gas depletion factors of $\mathcal{F}_{\rm gas} \sim 2$--5\% suggest the CCKB formed during the last few percent of the gas disk's life. In the photoevaporative disk models of \citet{Kunitomo2021}, the last vestiges of the outer gas disk are eliminated on timescales $< 10^5$ yr (see their Fig.~15). Note that sufficient disk gas is crucial for forming CCKBOs, not just because the streaming instability relies on mutual gas-dust drag, but also because lowering $\mathcal{F}_{\rm gas}$ while keeping the particle mass fixed makes super-Roche densities and gravitational instability that much harder to attain.

A related problem is why the CCKB truncates at 47 au. Despite deeper Kuiper belt surveys that have discovered objects in increasing abundance beyond this distance, the CCKB appears to retain its identity as a radially narrow and flat ring (see e.g.~Fig.~4 of \citealt{Fraser2024}). As explained above, we ruled out forming the CCKB inside a strict dust trap from a large-amplitude gas bump. However, a smaller-amplitude gas bump, superposed on a smooth gas disk having an outwardly decreasing pressure profile, can slow (but not stop) the inward drift of particles as they exit the bump. This slow down may promote planetesimal formation locally (cf.~\citealt{Carrera2021,Carrera2022}). Thus a gas bump could still be implicated in the formation of the CCKB, and explain the belt's radial localization, in a hybrid of the dust trapping and drift-and-form scenarios. There is a web of giant planet secular resonances inside 42 au (e.g.~\citealt{Duncan1995}) that truncates the CCKB at its inner edge. Perhaps these resonances also perturbed the CCKB's parent gas disk in a way that can help to explain the outer edge.

\section*{Acknowledgements}

We thank Lim Jeonghoon, Anders Johansen, J.J.~Kavelaars, Eve Lee, David Nesvorn\'{y}, Jake Simon, and Andrew Youdin for useful exchanges. 
\rla{We thank the referee for motivating substantive additions to our paper, in particular the zoomed-in simulations that recover retrograde spinning clumps.} 
R.L. acknowledges support from the Heising-Simons Foundation 51 Pegasi b Fellowship, and E.C. is grateful for a Simons Investigator grant and NSF grant 2205500 for ACCESS supercomputer allocations. This research used the Savio computational cluster resource provided by the Berkeley Research Computing program at the University of California, Berkeley (supported by the UC Berkeley Chancellor, Vice Chancellor for Research, and Chief Information Officer).

\added{\software{ATHENA \citep{Stone2008, Bai2010}, 
          Matplotlib \citep{Matplotlib}, 
          Numpy \& Scipy \citep{Numpy},
          Pyridoxine \citep{Pyridoxine},
          PLAN \citep{PLAN}.}}




\bibliographystyle{aasjournal}
\bibliography{refs}

\begin{thebibliography}{}
\expandafter\ifx\csname natexlab\endcsname\relax\def\natexlab#1{#1}\fi
\providecommand{\url}[1]{\href{#1}{#1}}
\providecommand{\dodoi}[1]{doi:~\href{http://doi.org/#1}{\nolinkurl{#1}}}
\providecommand{\doeprint}[1]{\href{http://ascl.net/#1}{\nolinkurl{http://ascl.net/#1}}}
\providecommand{\doarXiv}[1]{\href{https://arxiv.org/abs/#1}{\nolinkurl{https://arxiv.org/abs/#1}}}

\bibitem[{{Andrews} {et~al.}(2018){Andrews}, {Huang}, {P{\'e}rez}, {Isella}, {Dullemond}, {Kurtovic}, {Guzm{\'a}n}, {Carpenter}, {Wilner}, {Zhang}, {Zhu}, {Birnstiel}, {Bai}, {Benisty}, {Hughes}, {{\"O}berg}, \& {Ricci}}]{Andrews2018}
{Andrews}, S.~M., {Huang}, J., {P{\'e}rez}, L.~M., {et~al.} 2018, \apjl, 869, L41, \dodoi{10.3847/2041-8213/aaf741}

\bibitem[{{Bai} \& {Stone}(2010)}]{Bai2010}
{Bai}, X.-N., \& {Stone}, J.~M. 2010, \apjs, 190, 297, \dodoi{10.1088/0067-0049/190/2/297}

\bibitem[{{Batygin} {et~al.}(2020){Batygin}, {Adams}, {Batygin}, \& {Petigura}}]{Batygin2020}
{Batygin}, K., {Adams}, F.~C., {Batygin}, Y.~K., \& {Petigura}, E.~A. 2020, \aj, 159, 101, \dodoi{10.3847/1538-3881/ab665d}

\bibitem[{{Bierson} \& {Nimmo}(2019)}]{Bierson2019}
{Bierson}, C.~J., \& {Nimmo}, F. 2019, \icarus, 326, 10, \dodoi{10.1016/j.icarus.2019.01.027}

\bibitem[{{Brown}(2001)}]{Brown2001}
{Brown}, M.~E. 2001, \aj, 121, 2804, \dodoi{10.1086/320391}

\bibitem[{{Carrera} \& {Simon}(2022)}]{Carrera2022}
{Carrera}, D., \& {Simon}, J.~B. 2022, \apjl, 933, L10, \dodoi{10.3847/2041-8213/ac6b3e}

\bibitem[{{Carrera} {et~al.}(2021){Carrera}, {Simon}, {Li}, {Kretke}, \& {Klahr}}]{Carrera2021}
{Carrera}, D., {Simon}, J.~B., {Li}, R., {Kretke}, K.~A., \& {Klahr}, H. 2021, \aj, 161, 96, \dodoi{10.3847/1538-3881/abd4d9}

\bibitem[{Chiang \& Youdin(2010)}]{Chiang2010}
Chiang, E., \& Youdin, A. 2010, Annual Review of Earth and Planetary Sciences, 38, 493, \dodoi{10.1146/annurev-earth-040809-152513}

\bibitem[{{Dullemond} {et~al.}(2018){Dullemond}, {Birnstiel}, {Huang}, {Kurtovic}, {Andrews}, {Guzm{\'a}n}, {P{\'e}rez}, {Isella}, {Zhu}, {Benisty}, {Wilner}, {Bai}, {Carpenter}, {Zhang}, \& {Ricci}}]{Dullemond2018}
{Dullemond}, C.~P., {Birnstiel}, T., {Huang}, J., {et~al.} 2018, \apjl, 869, L46, \dodoi{10.3847/2041-8213/aaf742}

\bibitem[{{Duncan} {et~al.}(1995){Duncan}, {Levison}, \& {Budd}}]{Duncan1995}
{Duncan}, M.~J., {Levison}, H.~F., \& {Budd}, S.~M. 1995, \aj, 110, 3073, \dodoi{10.1086/117748}

\bibitem[{{Fraser} {et~al.}(2017){Fraser}, {Bannister}, {Pike}, {Marsset}, {Schwamb}, {Kavelaars}, {Lacerda}, {Nesvorn{\'y}}, {Volk}, {Delsanti}, {Benecchi}, {Lehner}, {Noll}, {Gladman}, {Petit}, {Gwyn}, {Chen}, {Wang}, {Alexand ersen}, {Burdullis}, {Sheppard}, \& {Trujillo}}]{Fraser2017}
{Fraser}, W.~C., {Bannister}, M.~T., {Pike}, R.~E., {et~al.} 2017, Nature Astronomy, 1, 0088, \dodoi{10.1038/s41550-017-0088}

\bibitem[{{Fraser} {et~al.}(2021){Fraser}, {Benecchi}, {Kavelaars}, {Marsset}, {Pike}, {Bannister}, {Schwamb}, {Volk}, {Nesvorny}, {Alexandersen}, {Chen}, {Gwyn}, {Lehner}, \& {Wang}}]{Fraser2021}
{Fraser}, W.~C., {Benecchi}, S.~D., {Kavelaars}, J.~J., {et~al.} 2021, \psj, 2, 90, \dodoi{10.3847/PSJ/abf04a}

\bibitem[{{Fraser} {et~al.}(2024){Fraser}, {Porter}, {Peltier}, {Kavelaars}, {Verbiscer}, {Buie}, {Stern}, {Spencer}, {Benecchi}, {Terai}, {Ito}, {Yoshida}, {Gerdes}, {Napier}, {Lin}, {Gwyn}, {Smotherman}, {Fabbro}, {Singer}, {Alexander}, {Arimatsu}, {Banks}, {Bray}, {Ramy El-Maarry}, {Ferrell}, {Fuse}, {Glass}, {Holt}, {Hong}, {Ishimaru}, {Johnson}, {Lauer}, {Leiva}, {S. Lykawka}, {Marschall}, {N{\'u}{\~n}ez}, {Postman}, {Quirico}, {Rhoden}, {Simpson}, {Schenk}, {Skrutskie}, {Steffl}, \& {Throop}}]{Fraser2024}
{Fraser}, W.~C., {Porter}, S.~B., {Peltier}, L., {et~al.} 2024, \psj, 5, 227, \dodoi{10.3847/PSJ/ad6f9e}

\bibitem[{{Gerbig} \& {Li}(2023)}]{Gerbig2023}
{Gerbig}, K., \& {Li}, R. 2023, \apj, 949, 81, \dodoi{10.3847/1538-4357/acca1a}

\bibitem[{{Gladman} \& {Volk}(2021)}]{Gladman_Volk2021}
{Gladman}, B., \& {Volk}, K. 2021, \araa, 59, 203, \dodoi{10.1146/annurev-astro-120920-010005}

\bibitem[{{Goldreich} \& {Tremaine}(1978)}]{GT1978}
{Goldreich}, P., \& {Tremaine}, S.~D. 1978, \icarus, 34, 227, \dodoi{10.1016/0019-1035(78)90164-1}

\bibitem[{{Gole} {et~al.}(2020){Gole}, {Simon}, {Li}, {Youdin}, \& {Armitage}}]{Gole2020}
{Gole}, D.~A., {Simon}, J.~B., {Li}, R., {Youdin}, A.~N., \& {Armitage}, P.~J. 2020, \apj, 904, 132, \dodoi{10.3847/1538-4357/abc334}

\bibitem[{{Gomes}(2021)}]{Gomes2021}
{Gomes}, R. 2021, \icarus, 357, 114121, \dodoi{10.1016/j.icarus.2020.114121}

\bibitem[{Grundy {et~al.}(2019)Grundy, Noll, Roe, Buie, Porter, Parker, Nesvorný, Levison, Benecchi, Stephens, \& Trujillo}]{Grundy2019}
Grundy, W., Noll, K., Roe, H., {et~al.} 2019, \icarus, \dodoi{https://doi.org/10.1016/j.icarus.2019.03.035}

\bibitem[{{Gulbis} {et~al.}(2006){Gulbis}, {Elliot}, \& {Kane}}]{Gulbis2006}
{Gulbis}, A. A.~S., {Elliot}, J.~L., \& {Kane}, J.~F. 2006, \icarus, 183, 168, \dodoi{10.1016/j.icarus.2006.01.021}

\bibitem[{{Hawley} {et~al.}(1995){Hawley}, {Gammie}, \& {Balbus}}]{Hawley1995}
{Hawley}, J.~F., {Gammie}, C.~F., \& {Balbus}, S.~A. 1995, \apj, 440, 742, \dodoi{10.1086/175311}

\bibitem[{{Holman} \& {Wisdom}(1993)}]{Holman_Wisdon_1993}
{Holman}, M.~J., \& {Wisdom}, J. 1993, \aj, 105, 1987, \dodoi{10.1086/116574}

\bibitem[{{Huang} {et~al.}(2018){Huang}, {Andrews}, {Dullemond}, {Isella}, {P{\'e}rez}, {Guzm{\'a}n}, {{\"O}berg}, {Zhu}, {Zhang}, {Bai}, {Benisty}, {Birnstiel}, {Carpenter}, {Hughes}, {Ricci}, {Weaver}, \& {Wilner}}]{Huang_DSHARP_2018}
{Huang}, J., {Andrews}, S.~M., {Dullemond}, C.~P., {et~al.} 2018, \apjl, 869, L42, \dodoi{10.3847/2041-8213/aaf740}

\bibitem[{{Huang} {et~al.}(2022){Huang}, {Gladman}, \& {Volk}}]{Huang2022}
{Huang}, Y., {Gladman}, B., \& {Volk}, K. 2022, \apjs, 259, 54, \dodoi{10.3847/1538-4365/ac559a}

\bibitem[{Hunter(2007)}]{Matplotlib}
Hunter, J.~D. 2007, Computing in Science Engineering, 9, 90, \dodoi{10.1109/MCSE.2007.55}

\bibitem[{{Jennings} \& {Chiang}(2021)}]{Jennings_Chiang2021}
{Jennings}, R.~M., \& {Chiang}, E. 2021, \mnras, 507, 5187, \dodoi{10.1093/mnras/stab2429}

\bibitem[{{Johansen} {et~al.}(2007){Johansen}, {Oishi}, {Mac Low}, {Klahr}, {Henning}, \& {Youdin}}]{Johansen2007a}
{Johansen}, A., {Oishi}, J.~S., {Mac Low}, M.-M., {et~al.} 2007, \nat, 448, 1022, \dodoi{10.1038/nature06086}

\bibitem[{{Johnston}(2019)}]{JohnBinaryTable}
{Johnston}, W.~R. 2019, {Binary Minor Planets Compilation V3.0}, NASA Planetary Data System, urn:nasa:pds:ast\_binary\_parameters\_compilation::3.0, \dodoi{10.26033/bb68-pw96}

\bibitem[{Jones {et~al.}(2001)Jones, Oliphant, Peterson, {et~al.}}]{Numpy}
Jones, E., Oliphant, T., Peterson, P., {et~al.} 2001, {SciPy}: Open source scientific tools for {Python}.
\newblock \url{http://www.scipy.org/}

\bibitem[{{Kavelaars} {et~al.}(2021){Kavelaars}, {Petit}, {Gladman}, {Bannister}, {Alexandersen}, {Chen}, {Gwyn}, \& {Volk}}]{Kavelaars2021}
{Kavelaars}, J.~J., {Petit}, J.-M., {Gladman}, B., {et~al.} 2021, \apjl, 920, L28, \dodoi{10.3847/2041-8213/ac2c72}

\bibitem[{{Klahr} \& {Schreiber}(2020)}]{Klahr_Schreiber_2020}
{Klahr}, H., \& {Schreiber}, A. 2020, \apj, 901, 54, \dodoi{10.3847/1538-4357/abac58}

\bibitem[{{Knezevic} {et~al.}(1991){Knezevic}, {Milani}, {Farinella}, {Froeschle}, \& {Froeschle}}]{Knezevic1991}
{Knezevic}, Z., {Milani}, A., {Farinella}, P., {Froeschle}, C., \& {Froeschle}, C. 1991, \icarus, 93, 316, \dodoi{10.1016/0019-1035(91)90215-F}

\bibitem[{{Kunitomo} {et~al.}(2021){Kunitomo}, {Ida}, {Takeuchi}, {Pani{\'c}}, {Miley}, \& {Suzuki}}]{Kunitomo2021}
{Kunitomo}, M., {Ida}, S., {Takeuchi}, T., {et~al.} 2021, \apj, 909, 109, \dodoi{10.3847/1538-4357/abdb2a}

\bibitem[{{Lau} {et~al.}(2025){Lau}, {Birnstiel}, {Stammler}, \& {Dr{\k{a}}{\.z}kowska}}]{Lau2025}
{Lau}, T. C.~H., {Birnstiel}, T., {Stammler}, S.~M., \& {Dr{\k{a}}{\.z}kowska}, J. 2025, arXiv e-prints, arXiv:2509.21437, \dodoi{10.48550/arXiv.2509.21437}

\bibitem[{{Lee}(2024)}]{Lee2024}
{Lee}, E.~J. 2024, \apjl, 970, L15, \dodoi{10.3847/2041-8213/ad5d8e}

\bibitem[{{Lee} {et~al.}(2022){Lee}, {Fuentes}, \& {Hopkins}}]{Lee2022}
{Lee}, E.~J., {Fuentes}, J.~R., \& {Hopkins}, P.~F. 2022, \apj, 937, 95, \dodoi{10.3847/1538-4357/ac8cfe}

\bibitem[{{Li}(2019)}]{PLAN}
{Li}, R. 2019, PLAN: A Clump-finder for Planetesimal Formation Simulations.
\newblock \doeprint{1911.001}

\bibitem[{Li(2020)}]{Pyridoxine}
Li, R. 2020, Pyridoxine: Handy Python Snippets for Athena Data.
\newblock \url{https://pypi.org/project/pyridoxine}

\bibitem[{{Li} \& {Youdin}(2021)}]{LY21}
{Li}, R., \& {Youdin}, A. 2021, arXiv e-prints, arXiv:2105.06042.
\newblock \doarXiv{2105.06042}

\bibitem[{{Li} {et~al.}(2018){Li}, {Youdin}, \& {Simon}}]{Li2018}
{Li}, R., {Youdin}, A.~N., \& {Simon}, J.~B. 2018, \apj, 862, 14, \dodoi{10.3847/1538-4357/aaca99}

\bibitem[{{Li} {et~al.}(2019){Li}, {Youdin}, \& {Simon}}]{Li2019}
---. 2019, \apj, 885, 69, \dodoi{10.3847/1538-4357/ab480d}

\bibitem[{{Lim} {et~al.}(2025){Lim}, {Simon}, {Li}, {Carrera}, {Baronett}, {Youdin}, {Lyra}, \& {Yang}}]{Lim2025}
{Lim}, J., {Simon}, J.~B., {Li}, R., {et~al.} 2025, \apj, 981, 160, \dodoi{10.3847/1538-4357/adb311}

\bibitem[{{Lim} {et~al.}(2024){Lim}, {Simon}, {Li}, {Armitage}, {Carrera}, {Lyra}, {Rea}, {Yang}, \& {Youdin}}]{Lim2024}
---. 2024, \apj, 969, 130, \dodoi{10.3847/1538-4357/ad47a2}

\bibitem[{{Lin} {et~al.}(2018){Lin}, {Lee}, \& {Chiang}}]{Lin2018}
{Lin}, J.~W., {Lee}, E.~J., \& {Chiang}, E. 2018, \mnras, 480, 4338, \dodoi{10.1093/mnras/sty2159}

\bibitem[{{Morbidelli} {et~al.}(2009){Morbidelli}, {Bottke}, {Nesvorn{\'y}}, \& {Levison}}]{Morbidelli2009}
{Morbidelli}, A., {Bottke}, W.~F., {Nesvorn{\'y}}, D., \& {Levison}, H.~F. 2009, \icarus, 204, 558, \dodoi{10.1016/j.icarus.2009.07.011}

\bibitem[{{Morbidelli} \& {Nesvorn{\'y}}(2020)}]{Morbidelli2020}
{Morbidelli}, A., \& {Nesvorn{\'y}}, D. 2020, Kuiper belt: formation and evolution, ed. D.~{Prialnik}, M.~A. {Barucci}, \& L.~{Young}, 25--59, \dodoi{10.1016/B978-0-12-816490-7.00002-3}

\bibitem[{{Nakagawa} {et~al.}(1986){Nakagawa}, {Sekiya}, \& {Hayashi}}]{Nakagawa1986}
{Nakagawa}, Y., {Sekiya}, M., \& {Hayashi}, C. 1986, \icarus, 67, 375, \dodoi{10.1016/0019-1035(86)90121-1}

\bibitem[{{Napier} {et~al.}(2023){Napier}, {Lin}, {Gerdes}, {Adams}, {Simpson}, {Porter}, {Weber}, {Markwardt}, {Gowman}, {Smotherman}, {Bernardinelli}, {Juri{\'c}}, {Connolly}, {Bryce Kalmbach}, {Portillo}, {Trilling}, {Strauss}, {Oldroyd}, {Trujillo}, {Chandler}, {Holman}, {Schlichting}, {McNeill}, \& {the DEEP Collaboration}}]{Napier2023}
{Napier}, K.~J., {Lin}, H.-W., {Gerdes}, D.~W., {et~al.} 2023, arXiv e-prints, arXiv:2309.09478, \dodoi{10.48550/arXiv.2309.09478}

\bibitem[{{Nesvorn{\'y}}(2015)}]{Nesvorny2015}
{Nesvorn{\'y}}, D. 2015, \aj, 150, 68, \dodoi{10.1088/0004-6256/150/3/68}

\bibitem[{{Nesvorn{\'y}} {et~al.}(2021){Nesvorn{\'y}}, {Li}, {Simon}, {Youdin}, {Richardson}, {Marschall}, \& {Grundy}}]{Nesvorny2021}
{Nesvorn{\'y}}, D., {Li}, R., {Simon}, J.~B., {et~al.} 2021, The Planetary Science Journal, 2, 27, \dodoi{10.3847/PSJ/abd858}

\bibitem[{{Nesvorn{\'y}} {et~al.}(2019){Nesvorn{\'y}}, {Li}, {Youdin}, {Simon}, \& {Grundy}}]{Nesvorny2019}
{Nesvorn{\'y}}, D., {Li}, R., {Youdin}, A.~N., {Simon}, J.~B., \& {Grundy}, W.~M. 2019, Nature Astronomy, 3, 808, \dodoi{10.1038/s41550-019-0806-z}

\bibitem[{{Nesvorn{\'y}} \& {Vokrouhlick{\'y}}(2019)}]{Nesvorny2019b}
{Nesvorn{\'y}}, D., \& {Vokrouhlick{\'y}}, D. 2019, \icarus, 331, 49, \dodoi{10.1016/j.icarus.2019.04.030}

\bibitem[{{Nesvorn{\'y}} {et~al.}(2022){Nesvorn{\'y}}, {Vokrouhlick{\'y}}, \& {Fraser}}]{Nesvorny2022}
{Nesvorn{\'y}}, D., {Vokrouhlick{\'y}}, D., \& {Fraser}, W.~C. 2022, \aj, 163, 137, \dodoi{10.3847/1538-3881/ac4bc9}

\bibitem[{{Noll} {et~al.}(2020){Noll}, {Grundy}, {Nesvorn{\'y}}, \& {Thirouin}}]{Noll2020}
{Noll}, K., {Grundy}, W.~M., {Nesvorn{\'y}}, D., \& {Thirouin}, A. 2020, Trans-Neptunian binaries (2018), ed. D.~{Prialnik}, M.~A. {Barucci}, \& L.~{Young}, 201--224, \dodoi{10.1016/B978-0-12-816490-7.00009-6}

\bibitem[{{Petit} {et~al.}(2023){Petit}, {Gladman}, {Kavelaars}, {Bannister}, {Alexandersen}, {Volk}, \& {Chen}}]{Petit2023}
{Petit}, J.-M., {Gladman}, B., {Kavelaars}, J.~J., {et~al.} 2023, \apjl, 947, L4, \dodoi{10.3847/2041-8213/acc525}

\bibitem[{{Poincar{\'e}}(1885)}]{Poincare1885}
{Poincar{\'e}}, H. 1885, Bulletin Astronomique, Serie I, 2, 109

\bibitem[{{Robinson} {et~al.}(2020){Robinson}, {Fraser}, {Fitzsimmons}, \& {Lacerda}}]{Robinson2020}
{Robinson}, J.~E., {Fraser}, W.~C., {Fitzsimmons}, A., \& {Lacerda}, P. 2020, \aap, 643, A55, \dodoi{10.1051/0004-6361/202037456}

\bibitem[{{Sch{\"a}fer} {et~al.}(2017){Sch{\"a}fer}, {Yang}, \& {Johansen}}]{Schafer2017}
{Sch{\"a}fer}, U., {Yang}, C.-C., \& {Johansen}, A. 2017, \aap, 597, A69, \dodoi{10.1051/0004-6361/201629561}

\bibitem[{{Schlichting} \& {Sari}(2008{\natexlab{a}})}]{Schlichting2008}
{Schlichting}, H.~E., \& {Sari}, R. 2008{\natexlab{a}}, \apj, 686, 741, \dodoi{10.1086/591073}

\bibitem[{{Schlichting} \& {Sari}(2008{\natexlab{b}})}]{Schlichting2008a}
---. 2008{\natexlab{b}}, \apj, 673, 1218, \dodoi{10.1086/524930}

\bibitem[{{Simon} {et~al.}(2016){Simon}, {Armitage}, {Li}, \& {Youdin}}]{Simon2016}
{Simon}, J.~B., {Armitage}, P.~J., {Li}, R., \& {Youdin}, A.~N. 2016, \apj, 822, 55, \dodoi{10.3847/0004-637X/822/1/55}

\bibitem[{{Simon} {et~al.}(2017){Simon}, {Armitage}, {Youdin}, \& {Li}}]{Simon2017}
{Simon}, J.~B., {Armitage}, P.~J., {Youdin}, A.~N., \& {Li}, R. 2017, \apjl, 847, \dodoi{10.3847/2041-8213/aa8c79}

\bibitem[{{Simon} {et~al.}(2024){Simon}, {Blum}, {Birnstiel}, \& {Nesvorn{\'y}}}]{Simon2024}
{Simon}, J.~B., {Blum}, J., {Birnstiel}, T., \& {Nesvorn{\'y}}, D. 2024, in Comets III, ed. K.~J. {Meech}, M.~R. {Combi}, D.~{Bockel{\'e}e-Morvan}, S.~N. {Raymodn}, \& M.~E. {Zolensky}, 63--94, \dodoi{10.2458/azu_uapress_9780816553631-ch003}

\bibitem[{{Simon} {et~al.}(2011){Simon}, {Hawley}, \& {Beckwith}}]{Simon2011}
{Simon}, J.~B., {Hawley}, J.~F., \& {Beckwith}, K. 2011, \apj, 730, 94, \dodoi{10.1088/0004-637X/730/2/94}

\bibitem[{{Stadler} {et~al.}(2025){Stadler}, {Benisty}, {Winter}, {Izquierdo}, {Longarini}, {Galloway-Sprietsma}, {Curone}, {Andrews}, {Bae}, {Facchini}, {Rosotti}, {Teague}, {Barraza-Alfaro}, {Cataldi}, {Cuello}, {Czekala}, {Fasano}, {Flock}, {Fukagawa}, {Garg}, {Hall}, {Hammond}, {Hilder}, {Huang}, {Ilee}, {Kanagawa}, {Lesur}, {Lodato}, {Loomis}, {Menard}, {Orihara}, {Pinte}, {Price}, {Yen}, {Wafflard-Fernandez}, {Wilner}, {W{\"o}lfer}, {Yoshida}, \& {Zawadzki}}]{Stadler2025}
{Stadler}, J., {Benisty}, M., {Winter}, A.~J., {et~al.} 2025, \apjl, 984, L11, \dodoi{10.3847/2041-8213/adb152}

\bibitem[{{Stammler} {et~al.}(2019){Stammler}, {Dr{\k{a}}{\.z}kowska}, {Birnstiel}, {Klahr}, {Dullemond}, \& {Andrews}}]{Stammler2019}
{Stammler}, S.~M., {Dr{\k{a}}{\.z}kowska}, J., {Birnstiel}, T., {et~al.} 2019, \apjl, 884, L5, \dodoi{10.3847/2041-8213/ab4423}

\bibitem[{{Stone} {et~al.}(2008){Stone}, {Gardiner}, {Teuben}, {Hawley}, \& {Simon}}]{Stone2008}
{Stone}, J.~M., {Gardiner}, T.~A., {Teuben}, P., {Hawley}, J.~F., \& {Simon}, J.~B. 2008, \apjs, 178, 137, \dodoi{10.1086/588755}

\bibitem[{{Van Laerhoven} {et~al.}(2019){Van Laerhoven}, {Gladman}, {Volk}, {Kavelaars}, {Petit}, {Bannister}, {Alexandersen}, {Chen}, \& {Gwyn}}]{Van_Laerhoven2019}
{Van Laerhoven}, C., {Gladman}, B., {Volk}, K., {et~al.} 2019, \aj, 158, 49, \dodoi{10.3847/1538-3881/ab24e1}

\bibitem[{{Xu} \& {Bai}(2022)}]{Xu_Bai_2022}
{Xu}, Z., \& {Bai}, X.-N. 2022, \apjl, 937, L4, \dodoi{10.3847/2041-8213/ac8dff}

\bibitem[{{Youdin} \& {Goodman}(2005)}]{Youdin_Goodman_2005}
{Youdin}, A.~N., \& {Goodman}, J. 2005, \apj, 620, 459, \dodoi{10.1086/426895}

\bibitem[{{Youdin} \& {Lithwick}(2007)}]{Youdin2007}
{Youdin}, A.~N., \& {Lithwick}, Y. 2007, \icarus, 192, 588, \dodoi{10.1016/j.icarus.2007.07.012}

\bibitem[{{Zagaria} {et~al.}(2023){Zagaria}, {Clarke}, {Booth}, {Facchini}, \& {Rosotti}}]{Zagaria2023}
{Zagaria}, F., {Clarke}, C.~J., {Booth}, R.~A., {Facchini}, S., \& {Rosotti}, G.~P. 2023, arXiv e-prints, arXiv:2311.08950, \dodoi{10.48550/arXiv.2311.08950}

\bibitem[{{Zhao} {et~al.}(2025){Zhao}, {Lau}, {Birnstiel}, {Stammler}, \& {Dr{\k{a}}{\.z}kowska}}]{Zhao2025}
{Zhao}, H., {Lau}, T. C.~H., {Birnstiel}, T., {Stammler}, S.~M., \& {Dr{\k{a}}{\.z}kowska}, J. 2025, \aap, 694, A205, \dodoi{10.1051/0004-6361/202452941}

\end{thebibliography}



\appendix

\section{Smoothed Gas Profiles near Radial Boundaries}
\label{app:smooth}

To model a limited radial portion of a pressure bump in the simulation domain (to better resolve the narrower central dust ring), both the gas density and gas azimuthal velocity need to be smoothed near the radial boundaries to avoid discontinuities. The procedure is  analogous to how the stellar vertical gravitational potential is handled in shearing-box simulations with periodic vertical boundary conditions.  The smoothing near the box edges should not affect our results near the box center. Specifically, we require 
\begin{equation}
    \left. \frac{\partial \rho_{\rm g}}{\partial x} \right|_{x=|L_X/2|} = 0
\end{equation}
at the radial boundaries so that the azimuthal gas velocity $u_{y}|_{x=|L_X/2|} = 0$.  We  modify the pure Gaussian  $\mathcal{G}(x)$ (used for either $\rho_{\rm g}$ or $\Sigma_{\rm g}$) to 
\begin{align}
    \mathcal{G}_{\rm s}(x) = \mathcal{G}(x) \left[1 - \mathcal{Q}(s)\right] + \mathcal{G}_{\rm b} \mathcal{Q}(s),
\end{align}
where 
\begin{align}
    s &= \frac{|x| - (\frac{1}{2}L_X - L_{\rm g})}{L_{\rm g}} \in [0,1] \\
\end{align}
is a normalized taper coordinate for ghost zone length $L_{\rm g}$, and 
\begin{align}
    Q(s) &= 
        \begin{cases}
        0, & s \le 0,\\
        6s^5 \;-\; 15s^4 \;+\; 10s^3, & 0 < s < 1,\\
        1, & s \ge 1,
        \end{cases}
\end{align}
is a quintic $\mathcal{C}^2$ smooth-step function used to merge the interior Gaussian $\mathcal{G}(x)$ with its constant boundary value $\mathcal{G}_{\rm b}\equiv \mathcal{G}(L_X/2)$.  In this way, for $|x| < L_X/2 - L_g$, $\mathcal{Q}(s) = 0$ so that $\mathcal{G}_{\rm s}(x) = \mathcal{G}(x)$, and when $|x|$ runs from $L_X/2 - L_g$ to $L_X/2$, $\mathcal{Q}(s)$ varies from $0$ to $1$ and $\mathcal{G}_{\rm s}(x)$ transitions to $\mathcal{G}_{\rm b}$ at the boundary.

The gas azimuthal velocity for a bump in geostrophic balance is
\begin{equation}
    u_{y} = \frac{c_{\rm s}^2}{2\Omega_0} \frac{\partial \ln \rho_{\rm g}}{\partial x},
\end{equation}
which requires computing $\frac{\partial}{\partial x} \ln \left[1 + \mathcal{A}\mathcal{G}_{\rm s}(x)\right]$ (see Eq. \ref{eq:G_bump} which defines the Gaussian bump profile with its amplitude $\mathcal{A}$ and width $w$). We have
\begin{align}
    \frac{\partial \mathcal{G}(x)}{\partial x} &= - \frac{x}{w^2} \mathcal{G}(x), \\
    \frac{\partial \mathcal{Q}}{\partial x} &= \frac{\partial \mathcal{Q}}{\partial s}\,\frac{\textnormal{d} s}{\textnormal{d} x}
        = \begin{cases}
        30\,s^2\,(s-1)^2\,\dfrac{\operatorname{sign}(z)}{L_g}, & 0 < s < 1,\\[6pt]
        0, & \text{otherwise,}
        \end{cases} 
\end{align}
and
\begin{align}
    \frac{\partial}{\partial x} \ln \left[1 + \mathcal{A} \mathcal{G}_{\rm s}(x) \right] &= \frac{\mathcal{A}}{1 + \mathcal{A} \mathcal{G}_{\rm s}(x)} \mathcal{G}'_{\rm s}(x), \\
    &= \frac{\mathcal{A}}{1 + \mathcal{A} \mathcal{G}_{\rm s}(x)}
    \left\{ \left[1 - \mathcal{Q}(s)\right] \frac{\partial \mathcal{G}(x)}{\partial x} + \left[ \mathcal{G}_{\rm b} - \mathcal{G}(x) \right] \frac{\partial \mathcal{Q}}{\partial x} \right\}.
\end{align}

\end{document}